\documentclass[pdflatex,sn-mathphys]{sn-jnl}

\jyear{2023}%

\theoremstyle{thmstyleone}%

\theoremstyle{thmstyletwo}%

\theoremstyle{thmstylethree}%
\makeatletter \def\ps@pprintTitle{  \let\@oddhead\@empty  \let\@evenhead\@empty  \def\@oddfoot{\hfill\thepage}  \def\@evenfoot{\thepage\hfill}} \makeatother

\begin{document}

\title [Accepted on Jan 17, 2024 in the International Journal of Information Security] {MLSTL-WSN: Machine Learning-based Intrusion Detection using SMOTETomek in WSNs}

\author*[1]{Md. Alamin Talukder} \email{alamin.cse@iubat.edu}
\author[2]{Selina Sharmin} \email{selina@cse.jnu.ac.bd}
\author[3]{Md Ashraf Uddin} \email{ashraf.uddin@deakin.edu.au}
\author[2]{Md Manowarul Islam} \email{manowar@cse.jnu.ac.bd}
\author[3]{Sunil Aryal}\email{sunil.aryal@deakin.edu.au}

\affil[1]{Department of Computer Science and Engineering, International University of Business Agriculture and Technology, Dhaka, Bangladesh}
\affil[2]{Department of Computer Science and Engineering, Jagannath University, Dhaka, Bangladesh}
\affil[3]{School of Information Technology, Deakin University, Waurn Ponds Campus, Geelong, Australia}

\abstract{
In the domain of cyber-physical systems, Wireless Sensor Networks (WSNs) play a pivotal role as infrastructures, encompassing both stationary and mobile sensors. These sensors self-organize and establish multi-hop connections for communication, collectively sensing, gathering, processing, and transmitting data about their surroundings. Despite their significance, WSNs face rapid and detrimental attacks that can disrupt functionality. Existing intrusion detection methods for WSNs encounter challenges such as low detection rates, computational overhead, and false alarms. These issues stem from sensor node resource constraints, data redundancy, and high correlation within the network.
To address these challenges, we propose an innovative intrusion detection approach that integrates Machine Learning (ML) techniques with the Synthetic Minority Oversampling Technique Tomek Link (SMOTE-TomekLink) algorithm. This blend synthesizes minority instances and eliminates Tomek links, resulting in a balanced dataset that significantly enhances detection accuracy in WSNs. Additionally, we incorporate feature scaling through standardization to render input features consistent and scalable, facilitating more precise training and detection. To counteract imbalanced WSN datasets, we employ the SMOTE-Tomek resampling technique, mitigating overfitting and underfitting issues. Our comprehensive evaluation, using the Wireless Sensor Network Dataset (WSN-DS) containing 374,661 records, identifies the optimal model for intrusion detection in WSNs.
The standout outcome of our research is the remarkable performance of our model. In binary classification scenarios, it achieves an accuracy rate of 99.78\%, and in multiclass classification scenarios, it attains an exceptional accuracy rate of 99.92\%. These findings underscore the efficiency and superiority of our proposal in the context of WSN intrusion detection, showcasing its effectiveness in detecting and mitigating intrusions in WSNs.
}


\maketitle

\section{Introduction}
\label{introduction}

Wireless Sensor Networks (WSNs) have emerged as a transformative technology that enables the collection, processing, and transmission of data from distributed sensor nodes. These nodes are equipped with various sensors and communication capabilities, allowing them to monitor and sense the environment \cite{gebremariam2023design}. WSNs find applications in diverse domains such as environmental monitoring, healthcare, smart cities, and industrial automation. They offer the advantage of remote and real-time data acquisition from inaccessible or hazardous locations, enabling efficient data-driven decision-making processes \cite{chataut2023unleashing, talukder2024machine}.
WSN security is of paramount importance due to the sensitive nature of the data transmitted and the potential vulnerabilities in the network \cite{yakubu2023resource}. WSNs face various security threats, including unauthorized access, data tampering, and denial of service attacks \cite{nimbalkar2023security}. The distributed and wireless nature of WSNs makes them more susceptible to these threats. Ensuring the confidentiality, integrity, and availability of data within WSNs is crucial to maintaining the trust and reliability of these networks \cite{alghamdi2023cascaded}.

Intrusion detection is a critical component of WSN security, aimed at detecting and mitigating malicious activities within the network \cite{heidari2022internet}. Traditional rule-based intrusion detection systems often rely on predefined signatures or thresholds, which are not effective in detecting sophisticated attacks that evolve \cite{sezgin2023aid4i}. Machine Learning (ML) techniques have emerged as a promising approach for WSN intrusion detection \cite{gebremariam2023design}. ML algorithms can learn from historical data and identify anomalies or patterns indicative of potential intrusions, enabling proactive and adaptive security measures \cite{talukder2023dependable, talukder2024machine}.
Machine Learning techniques in WSN enable the development of intelligent intrusion detection systems \cite{ghazal2022data}. These systems can analyze vast collected data, identify abnormal patterns, and distinguish between normal and malicious behavior \cite{talukder2022machine}. ML algorithms, such as decision trees, random forests, neural networks, and gradient boosting methods, can extract valuable insights from complex WSN datasets, improving the accuracy and effectiveness of intrusion detection mechanisms \cite{talukder2023dependable, gebremariam2023design}.

Existing works in WSN intrusion detection face several challenges. Limited scalability is a significant concern, as WSNs often involve a large number of sensor nodes, leading to increased computational complexity \cite{sharmin2023energy}. High false positive rates and inadequate adaptability to evolving attack techniques are also challenges that need to be addressed \cite{tan2019wireless, ifzarne2021anomaly, alruhaily2021multi}. Additionally, the imbalanced nature of WSN datasets, with a majority of normal instances and a small number of intrusion instances, poses challenges for achieving accurate detection results \cite{singh2020fuzzy, chandre2022intrusion, putrada2022xgboost}.
Our proposed model, which combines Machine Learning (ML) techniques with the Synthetic Minority Oversampling Technique Tomek (SMOTE-Tomek), offers a significant advancement in intrusion detection for WSNs (WSNs) \cite{tan2019wireless, zhang2020effective, mohammadi2023novel}. This model holds great importance in addressing the limitations of existing approaches and providing an optimal solution for intrusion detection in WSNs.

The primary motivation behind our work stems from the inherent imbalance in WSN intrusion detection datasets, with a majority of normal instances and a small number of intrusion instances \cite{putrada2022xgboost}. This imbalance can compromise the accuracy of detection results. Our model aims to overcome this challenge by employing the SMOTE-Tomek technique, synthesizing minority instances and removing Tomek links to achieve a balanced dataset \cite{chandra2023median}. This ensures more accurate and reliable intrusion detection by enhancing the representation of both normal and intrusion instances. Additionally, our model harnesses the power of various ML algorithms, including Decision Trees (DT), Random Forests (RF), Multilayer Perception (MLP), K-Nearest Neighbor (KNN), and gradient boosting algorithms such as Extreme boosting (XGB) and Light gradient boosting (LGB). These algorithms enable the development of a robust intrusion detection system capable of learning complex patterns and anomalies from WSN data. The adaptability of ML techniques allows our model to continuously learn and update its knowledge to effectively detect evolving attack techniques.

The significance of our model lies in its proactive and adaptive security measures for WSNs. By accurately identifying and mitigating intrusions in real-time, our model safeguards the integrity and privacy of transmitted data. This capability is crucial in domains like healthcare, industrial control systems, and environmental monitoring, where data security and integrity are paramount.

This paper makes the following key contributions:

\begin{itemize}
    \item Innovative Intrusion Detection Approach: Introducing a novel intrusion detection approach that integrates ML techniques with the SMOTE-Tomek algorithm. By synthesizing minority instances and removing Tomek links, our model achieves a balanced dataset, resulting in improved accuracy and reliability in WSNs.

    \item Feature Scaling for Enhanced Effectiveness: Employing feature scaling through standardization to transform input features into a consistent and scalable range. This facilitates more accurate training and detection of intrusions. Additionally, applying the SMOTE-Tomek resampling technique addresses the imbalanced nature of WSN datasets, mitigating overfitting and underfitting issues.

    \item Comprehensive Model Evaluation: Conducting a thorough evaluation of our ML-based intrusion detection model using various performance metrics, including accuracy, precision, recall, and F1 score. Through extensive analysis, we identify the best-performing model for detecting intrusions in WSNs.

    \item Performance Comparison: Comparing the performance of our model with existing approaches to demonstrate its effectiveness and superiority in detecting and mitigating intrusions in WSNs.
    
\end{itemize}

Overall, our contributions enhance the field of intrusion detection in WSNs by introducing a novel machine learning-based approach that addresses the challenges of imbalanced datasets. The proposed work combined with effective feature scaling and resampling techniques, provides an accurate and reliable solution for detecting intrusions in real-time. The extensive evaluation and performance analysis validate the effectiveness of our model, positioning it as a valuable tool for enhancing the security of WSNs.

The remaining sections of this paper are structured as follows:
In Section \ref{sec:Related}, we delve into a comprehensive review of the existing literature, with a specific focus on intrusion detection on WSNs. Section \ref{sec:Method} provides a detailed account of our research methodology, including an extensive description of the dataset employed.
The experimental setup and the ensuing performance evaluation are elaborated upon in Section \ref{sec:Results}.
Subsequently, Section \ref{sec:Discussion} is dedicated to an exhaustive analysis and discussion of our proposed research.
Finally, Section \ref{sec:Conclusion} encapsulates the concluding remarks and underscores potential avenues for future research.

\section{Related Works}
\label{sec:Related}

Intrusion detection in wireless sensor networks (WSNs) has received significant attention in recent years due to the increasing number of attacks on WSNs. Various machine learning (ML)-based intrusion detection models had been proposed in the literature to address the limitations of traditional intrusion detection methods in WSNs. 

\cite{tan2019wireless} introduced an approach that employs the Synthetic Minority Oversampling Technique (SMOTE) to address dataset imbalance, followed by training a classifier for intrusion detection using the Random Forest algorithm. Simulations were performed on a standard intrusion dataset, demonstrating that the Random Forest algorithm achieved an accuracy of 92.39\%, surpassing other algorithms in comparison. Furthermore, by applying SMOTE to oversample the minority samples, the accuracy of the Random Forest classifier improved to 92.57\%. This indicates that the proposed method effectively addresses class imbalance issues and enhances intrusion detection performance.

\cite{rezvi2021data} presented a data mining approach to detect various types of DoS attacks, where they applied several classification algorithms, including KNN, Naïve Bayes, Logistic Regression, Support Vector Machine (SVM), and Artificial Neural Network (ANN), to the dataset and assessed their performance in identifying these attacks. The analysis revealed that ANN achieved the highest accuracy at 98.56\%, followed closely by KNN at 98.4\%. These results suggest that ANN and KNN are strong candidates for intrusion detection, making them suitable recommendations for network specialists and analysts due to their effectiveness in detecting and predicting such attacks compared to the other algorithms tested.

\cite{meng2022novel} presented a LightGBM-based intrusion detection method for resource-constrained WSNs. It employed the SMOTE-Tomek technique for dataset balancing and feature selection using an iterative LightGBM tree model, SHAP analysis, and Recursive Feature Elimination. Model parameters were optimized with the Optuna algorithm. Compared to standard methods, this approach achieved exceptional performance, with detection rates exceeding 99\% for all attack types and a 46\% reduction in modeling time due to feature dimension reduction.

\cite{singh2020fuzzy} developed a fuzzy rule-based system for intrusion prevention in WSNs, involving three phases: feature extraction, membership value computation, and fuzzy rule application. Nodes were categorized as "red" (malicious, blocked), "orange" (potentially malicious, marked suspicious), and "green" (non-malicious, allowed). The system considered parameters like packet transmission, energy usage, signal strength, received packets, and Packet Delivery Ratio (PDR). Evaluation yielded a 98.29\% accuracy, surpassing other fuzzy rule-based systems. The key advantage was its ability to block malicious nodes and prevent intrusions.

\cite{alruhaily2021multi} proposed a multi-tier intrusion detection framework for WSNs, implementing a defense-in-depth security approach with two detection layers. In the initial layer, situated at the distributed network edge sensors, real-time decision-making for inspected packets was accomplished using a Naive Bayes classifier. The second layer, positioned in the cloud, employed a Random Forest multi-class classifier for thorough packet analysis. The outcomes demonstrated that their multi-layer detection model achieved impressive performance scores, including a 100\% precision rate for Normal attacks, 90.4\% for Flooding attacks, 99.5\% for Scheduling attacks, 97\% for Grayhole attacks, and 99.9\% for Blackhole attacks.

To offer intrusion prevention methods employing deep packet inspection based on deep learning techniques.  \cite{chandre2022intrusion} introduced a deep learning model that utilized a convolutional neural network. This model encompassed two essential phases: detecting intrusions and preventing them. It acquired meaningful feature representations from a sizable labeled dataset and performed accurate classifications.  They harnessed the power of the convolutional neural network to prevent intrusions in WSNs. They employed the WSN-DS to assess the system's effectiveness. The test outcomes indicated that the proposed system achieved a remarkable accuracy rate of 97\%, surpassing existing solutions. Their work could serve as a reference point for future research in deep learning and intrusion prevention.

\cite{dener2022stlgbm} developed STLGBM-DDS, an ensemble intrusion detection system on the Apache Spark platform in Google Colab. It combined LightGBM with data balancing (using SMOTE and Tomek-Links, STL) and feature selection (Information Gain Ratio). The study evaluated the impact of these stages on system performance using various parameters. The proposed method achieved an outstanding overall accuracy of 99.95\%. Specifically, it achieved impressive accuracy rates: 99.99\% for Normal, 99.96\% for Grayhole, 99.98\% for Blackhole, 99.92\% for TDMA, and 99.87\% for Flooding classes. These results demonstrated the system's exceptional success in detecting Denial of Service (DoS) attacks in WSNs compared to existing methods.

\cite{ifzarne2021anomaly} developed an intrusion detection model tailored to WSNs characteristics, using information gain ratio (IGR) and the online Passive Aggressive (PA) classifier. Experiments on a WSN-DS dataset resulted in the proposed model achieving a 96\% detection rate for normal behavior or attacks. Detection accuracies were 86\% for scheduling, 68\% for grayhole, 63\% for flooding, and 46\% for blackhole attacks, with 99\% accuracy for normal traffic. These findings suggest that offline learning-based models can effectively detect anomalies in WSNs, potentially replacing online learning in some cases.

An optimized collaborative intrusion detection system (OCIDS) was developed by \cite{elsaid2020optimized} for WSNs using an enhanced artificial bee colony optimization (BCO) algorithm. It improved the accuracy of intrusion detection and resource efficiency. The OCIDS also enhanced the weighted support vector machine (SVM) algorithm to reduce false alarms. Collaboration among sensor nodes, cluster heads, and the Base Station (BS) improved intrusion detection. In tests, the OCIDS outperformed other systems with a 97.9\% detection rate and a 1.8\% false alarm rate, demonstrating a clear advantage.

\cite{putrada2022xgboost} explored the application of XGBoost in Intrusion Detection Systems (IDS) for imbalanced data in WSNs cyber-attacks. To assess its performance, they compared XGBoost to decision trees and naive Bayes, employing various evaluation metrics. The results showed XGBoost's superiority, with the highest AUC values across scheduling, normal, grayhole, flooding, and blackhole classes, achieving 98.7\%, 99.63\%, 99.94\%, 99.97\%, and 99.99\%, respectively. This research contributes to the advancement of IDS for imbalanced WSN data, offering valuable insights for further investigations in this domain.

\cite{jiang2020slgbm} proposed Sequence-LightGBM (SLGBM) as an intrusion detection method for WSNs. Initially, the Sequence Backward Selection (SBS) algorithm was applied to reduce data dimensionality in the feature space of the original traffic data, thus reducing computational overhead. Subsequently, the LightGBM algorithm was utilized to detect various network attacks. Experimental results based on the WSN-DS dataset showed that the F-measure of our proposed SLGBM was significantly superior to current typical detection methods, achieving scores of 99.8\% for Normal, 99.4\% for Blackhole, 99.1\% for Grayhole, 96.5\% for Flooding, and 96.1\% for Scheduling attack detections, respectively.

\cite{khan2022enhancing} presented a real-time Intrusion Detection System (IDS) based on a deep autoencoder for distinguishing malicious actions within IIoT-driven IICS networks. The model was formulated using a Long Short-Term Memory (LSTM) autoencoder design specifically tailored to identify invasive events in IICS networks. The experimental results, conducted on two benchmark datasets, namely the gas pipeline and UNSW-NB15 datasets, demonstrated the superior performance of the proposed IDS when compared to other compelling models. The proposed model achieved an accuracy rate of 97.95\% for the gas pipeline data and 97.62\% for the UNSW-NB15 dataset, showcasing its effectiveness in intrusion detection within industrial control systems.

\cite{khan2022federated} proposed the federated-SRUs IDS, a novel model for IoT-based ICS security. It employs an improved SRUs architecture to address computational issues and uses federated learning for collaborative, privacy-preserving model building. Experimental validation on gas pipeline-based ICS data demonstrates real-time intrusion detection with superior performance compared to existing approaches. The federated-SRUs IDS emerges as an efficient method for IoT-based ICS network security.

\cite{khan2021new} proposed an autoencoder-based framework using convolutional and recurrent networks for cyber threat detection in IIoT networks. The two-step sliding window enhances feature learning, transforming malicious points into fixed-length series. Fully connected networks leverage extracted features for attack event classification and explanation. Empirical results highlight the framework's efficacy, outperforming contemporary methods and showcasing suitability for real-world IIoT networks.

\cite{khan2024fed} proposed Fed-Inforce-Fusion, a privacy-preserving FL-based IDS for cyber-attack detection in IoMT networks. The model employs reinforcement learning to capture latent relationships in medical data. In a distributed FL system, Secure Healthcare Systems nodes collaboratively train the IDS model, ensuring privacy. A fusion/aggregation strategy dynamically involves clients, enhancing model performance and reducing communication overhead. Theoretical and experimental analyses confirm Fed-Inforce-Fusion's superiority in detecting complex attack vectors, establishing its efficacy for real-world IoMT networks.

\cite{ravindra2023etelmad} proposed an Enhanced Transient Extreme Learning Machine Anomaly Detection method to address sensor data anomalies. The process involved data compression, prediction using an optimized Extreme Learning Machine (ELM) with Enhanced Transient Search Arithmetic Optimization (ETSAO), and anomalous detection through dynamic thresholding. Data preprocessing eliminated duplicates, and Piecewise Aggregate Approximation efficiently compressed the data. This approach achieved a low-dimensional feature set, reducing computation/training time in WSN environments. The architecture attained an impressive 96.90\% accuracy on the WSN-DS dataset, surpassing other approaches. Simulation on the PYTHON platform confirmed the efficacy of the proposed anomaly detection method.

\cite{alruwaili2023red} introduced the RKOA-AEID (Red Kite Optimization Algorithm with Average Ensemble Model for Intrusion Detection) for securing IoT-based WSNs. The methodology includes pre-processing with min-max normalization, RKOA-based feature selection, and an average ensemble learning model for intrusion detection. Additionally, the Lévy-fight chaotic whale optimization Algorithm (LCWOA) optimizes hyperparameters for the ensemble models. The RKOA-AEID algorithm was evaluated on the benchmark WSN-DS dataset, achieving an improved accuracy of 98.94\% compared to other approaches.

\cite{moundounga2024stochastic} proposed an innovative Stochastic ML-Based Attack Detection System for WSNs, combining Hidden Markov Models (HMMs) and Gaussian Mixture Models (GMMs). The system utilized Principal Component Analysis for dimensionality reduction in WSN datasets, preserving vital routing features while reducing variables. Iterative machine learning Expectation-Maximization was employed to train HMMs and GMMs, enabling accurate detection and classification of malicious activities and routing errors. Experimental results showed that a configuration of 3 HMMs and 4 GMMs achieved an outstanding accuracy of 94.55\%, highlighting the system's superior performance. This approach presented a promising solution to enhance WSN security.


Here, the related works highlight the importance of intrusion detection in WSNs and the ongoing efforts to develop effective machine learning-based intrusion detection models. Our proposed hybrid machine learning model offers a promising solution for intrusion detection in WSNs, providing improved accuracy and performance compared to previous methods.

\section{Methodology}
\label{sec:Method}
We integrated ML techniques with the SMOTE-Tomek algorithm to achieve a balanced dataset, resulting in improved intrusion detection for WSNs.The experiments conducted to validate the proposed intrusion detection approach in WSNs involved rigorous testing and evaluation steps. Below are detailed descriptions of the experimental procedures and implementation specifics:

\begin{itemize}
    
\item{Data Collection}
Raw data from WSNs was collected to form the foundation for the experimental dataset. The data encompassed diverse scenarios and network conditions to ensure the model's robustness and applicability across real-world scenarios.

\item{Preprocessing Techniques}
A crucial aspect of the methodology involved the preprocessing of the collected data. Standardization was applied to normalize input features, ensuring a consistent scale for all variables. Additionally, label encoding was employed to convert categorical target features into a numerical format, facilitating compatibility with machine learning algorithms.

\item{Data Balancing}
To address the challenge of imbalanced datasets and potential overfitting, the Synthetic Minority Over-sampling Technique (SMOTE) combined with the Tomek links removal method (SMOTETomek) was utilized. This technique ensured a balanced representation of both normal and intrusion instances in the dataset, contributing to improved model generalization.

\item{Data Splitting}
The experimental dataset underwent k-fold cross-validation, specifically with 10 folds, to split it into training and testing sets. This iterative splitting process helps in assessing the model's performance across various subsets of the data, promoting robustness and generalization.

\item{Model Building}
Several machine learning algorithms were implemented for intrusion detection model development. The ensemble method Random Forest (RF), decision tree (DT), multilayer perceptron (MLP), k-nearest neighbors (KNN), XGBoost (XGB), and LightGBM (LGB) were chosen for their diverse strengths in capturing different patterns within the data.

\item{Model Evaluation}
The performance of each developed model was meticulously evaluated using the test dataset. Various evaluation metrics, including accuracy, precision, recall, and F1-score, were employed to provide a comprehensive understanding of the models' effectiveness and reliability in detecting and classifying intrusions.

\item{Model Selection}
The final step involved selecting the best-performing model based on the comprehensive evaluation results. The model exhibiting superior performance across multiple metrics was deemed optimal for intrusion detection in WSNs. This selection process ensures the chosen model's suitability for deployment in real-world scenarios.

\end{itemize}

\begin{figure*}[!htbp]
	\centering
	{\includegraphics[scale=.20]{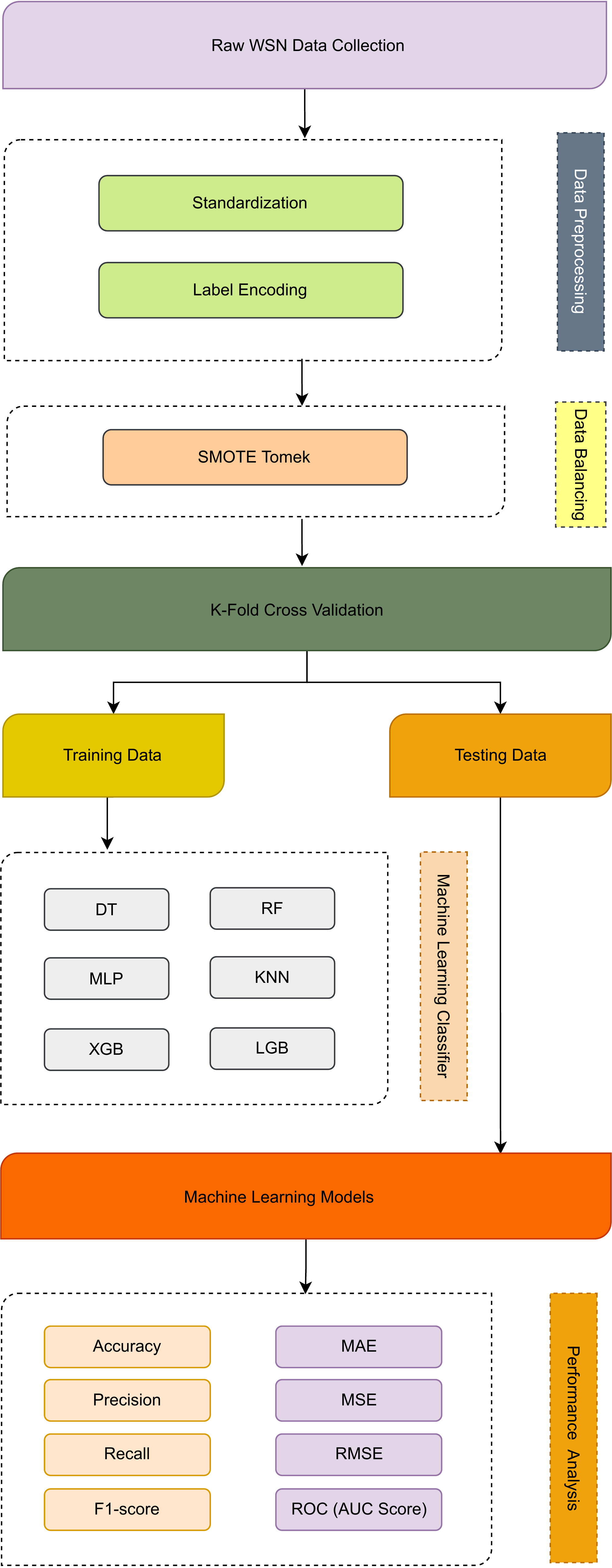}}
	\caption{The proposed intrusion detection approach in wireless sensor network}
	\label{fig:proposal}
\end{figure*}

The proposed intrusion detection approach, as illustrated in Figure \ref{fig:proposal}, integrates these experimental steps to create a robust and effective solution for enhancing the security of Wireless Sensor Networks. The detailed implementation and evaluation procedures guarantee the reliability and practicality of the developed intrusion detection models.

The overall algorithmic steps of our proposal for intrusion detection in WSNs is shown in Algorithm \ref{Algo:a1} as follows:

\begin{algorithm}
  \caption{Intrusion Detection In WSNs Using ML}
  \label{Algo:a1}
  \begin{algorithmic}[1]
    \Procedure{IntrusionDetectionWSN}{$file\_path, target\_column$}
      \State \textbf{Input:} WSN-DS as a CSV file path (\textit{file\_path})
      \State \textbf{Output:} Trained ML models, Evaluate Performance of each model
      \State

      \Procedure{ReadCSV}{$file\_path$}
        \State Initialize DataFrame: $df$
        \State $df \gets$ Read CSV file located at $file\_path$
      \EndProcedure
      \State

      \Procedure{PreprocessData}{$df, target\_column$}
        \State Initialize Features: $X$
        \State Initialize Target Variable: $y$
        \State

        \State $X, y \gets$ Extract features and target variable from $df$
        \State

        \State \Call{ApplyStandardization}{$X$}
          \State $X \gets$ Standardize numerical features in $X$
        \State

        \State \Call{EncodeLabels}{$y$}
          \State $y \gets$ Encode categorical labels in $y$ using LabelEncoder
        \EndProcedure
        \State

    \Procedure{DataBalancing}{$X, y$}
        \State \Call{BalanceData}{$X, y$}
          \State $X, y \gets$ Apply SMOTETomek to balance the dataset
        \EndProcedure
      \State

      \Procedure{SplitData}{$X, y$}
        \State Initialize K-Fold Cross-Validator: $kf$
        \State $kf \gets$ Initialize KFold(n\_splits=10, shuffle=True)
        \State

        \For{$train\_index, test\_index$ \textbf{in} $kf.split(X,y)$}
          \State $X\_train, X\_test \gets$ Split $X$ into training and testing sets
          \State $y\_train, y\_test \gets$ Split $y$ into training and testing sets
          \State

          \State \Call{BuildModels}{$X\_train, y\_train$}
            \State Initialize Each ML Model: $DT$, $RF$, $MLP$, $KNN$, $LGB$, $XGB$
            \State Train Each ML Model 
            \State

          \State \Call{EvaluatePerformance}{$X\_test, y\_test$}
            \State Evaluate Various Performances For Each ML Model
        \EndFor
    \EndProcedure
   \EndProcedure

  \end{algorithmic}
\end{algorithm}
\subsection{Dataset Description}

The Wireless Sensor Network DoS Detection Dataset (WSN-DS) \cite{almomani2016wsn} is a comprehensive collection crafted for the specific purpose of identifying Denial-of-Service (DoS) attacks in WSNs (WSNs). This dataset, consisting of 374,661 records, was curated using the LEACH protocol, a widely adopted routing protocol in WSNs, and includes instances of four types of DoS attacks (Blackhole, Grayhole, Flooding, and Scheduling) as well as normal network behavior scenarios. Data for WSN-DS was acquired using the Network Simulator 2 (NS-2) \cite{tv2006network}, processed to produce 23 features, enriching the dataset with valuable insights a well-established network simulation tool, and the dataset's performance evaluation and analysis are facilitated through integration with the Waikato Environment for Knowledge Analysis (WEKA) Toolbox, a versatile open-source data mining software suite developed at the University of Waikato in New Zealand. This resource empowers researchers to explore and combat DoS threats effectively within WSNs, enhancing network security in these environments \cite{hall2009weka, bouckaert2010weka}.

\subsection{Data Preprocessing}

Data preprocessing plays a crucial role in the success of any intrusion detection system in Wireless Sensor Networks (WSN). It involves several steps to ensure that the data is in a suitable format for machine learning models. In this section, we describe the key data preprocessing steps we have applied to our dataset.

\subsubsection{Standardization}

Standardization is the process of transforming data such that it has a mean of 0 and a standard deviation of 1. This step is essential when dealing with features that have different scales or units. In WSN intrusion detection, sensor data may have varying measurement units and scales. Standardization helps in bringing all features to a common scale, which is important for machine learning algorithms that rely on distance metrics or gradient-based optimization.

The formula for standardization is as follows:
\[X_{\text{standardized}} = \frac{X - \mu}{\sigma}\]

Where:
\begin{itemize}
\item $X_{\text{standardized}}$ is the standardized value of feature $X$.
\item $X$ is the original feature value.
\item $\mu$ is the mean of the feature $X$.
\item $\sigma$ is the standard deviation of the feature $X$.
\end{itemize}

By standardizing the data, we ensure that each feature contributes equally to the learning process, preventing some features from dominating others.

\subsubsection{Label Encoding}

Intrusion detection datasets often contain categorical variables, such as attack types or sensor identifiers, that need to be converted into numerical values for machine learning models. Label encoding is a technique used to transform categorical data into numerical labels.
\begin{itemize}
\item{Binary Classification}

For binary classification, where we have two classes, "Normal" and "Attack," we perform label encoding as shown in Table \ref{tab:binary-label-encoding}.

\begin{table}[htbp]
  \centering
  \caption{Label Encoding for Binary Classification}
  \label{tab:binary-label-encoding}
  \begin{tabular}{|c|c|}
    \hline
    Class  & Label Encoding \\
    \hline
    Normal & 0              \\
    Attack & 1              \\
    \hline
  \end{tabular}
\end{table}

\item{Multiclass Classification}

In the case of multiclass classification, where we have multiple classes, such as "Normal," "Grayhole," "Blackhole," "TDMA," and "Flooding," we perform label encoding as shown in Table \ref{tab:multiclass-label-encoding}.

\begin{table}[htbp]
  \centering
  \caption{Label Encoding for Multiclass Classification}
  \label{tab:multiclass-label-encoding}
  \begin{tabular}{|c|c|}
    \hline
    Class     & Label Encoding \\
    \hline
    Normal    & 0              \\
    Grayhole  & 1              \\
    Blackhole & 2              \\
    TDMA      & 3              \\
    Flooding  & 4              \\
    \hline
  \end{tabular}
\end{table}

After label encoding, the categorical features are represented as numerical values, making them suitable for use in machine learning algorithms.

Label encoding provides a straightforward way to convert categorical classes into a format that can be used as input for machine learning models.
\end{itemize}

In summary, data preprocessing is a critical step in preparing data for intrusion detection in WSNs. Standardization ensures that all features have a consistent scale, while label encoding converts categorical data into a numerical format, enabling the application of various machine learning algorithms to build an effective intrusion detection model.

\subsection{Data Balancing using SMOTE-TomekLink}

Data balancing refers to the process of adjusting the class distribution within a dataset to ensure that each class or category has a similar number of samples or observations. This is particularly important when dealing with classification tasks where the classes are imbalanced, meaning that one or more classes have significantly more or fewer instances than others. In the context of data balancing for Wireless Sensor Networks (WSN), the WSN-DS dataset exhibits a significant class imbalance issue, impacting classification performance, especially in the minority classes, which negatively affects detection rates. Traditional imbalance mitigation methods, such as undersampling, significantly reduce the usable normal network traffic data, while oversampling techniques alone lead to excessive data size inflation and noise. To address this imbalance, we employed the SMOTE-TomekLink (Synthetic Minority Over-sampling Technique combined with Tomek links) method to balance our dataset. This approach combines SMOTE and Tomek-Links oversampling and undersampling methods.

SMOTE is an oversampling technique that generates synthetic data for the minority class by interpolating between existing samples and their neighbors, effectively increasing the number of minority class samples. This approach helps mitigate overfitting issues associated with random oversampling methods and has been widely adopted to address class imbalance problems. On the other hand, Tomek-Links is an undersampling technique designed to remove instances on the "Tomek link," which are pairs of data points from different classes that are close to each other in the dataset. Removing these pairs helps separate minority and majority classes, reducing noise in the majority class. By combining the SMOTE and Tomek-Links (STL), the study aims to effectively tackle the imbalanced class problem in WSN data, providing a balanced dataset for more accurate intrusion detection. This strategy enhances class separation, stabilizes data distribution, and ultimately improves the performance of intrusion detection systems in the context of WSNs. In the following, we present the data balancing results using SMOTE-Tomek-Link compared to the unbalanced dataset.

\begin{itemize}

\item{Binary Classification}

In the binary classification dataset, we had the following distribution of classes without and with SMOTE-TomekLink is shown in Table \ref{tab:binary-distribution}:

\begin{table}[!h]
  \centering
  \begin{tabular}{lll}
    \hline
    Class  & WoSTL & WiSTL \\
    \hline
    Normal & 340,066 & 340,056 \\
    Attack & 34,595 & 339,610 \\
    \hline
  \end{tabular}
  \caption{Binary Classification Class Distribution without and with SMOTE-TomekLink}
  \label{tab:binary-distribution}

\end{table}


The STL technique has effectively balanced the dataset, resulting in a nearly equal number of samples for both the "Normal" and "Attack" classes. This balanced dataset (WiSTL) is better suited for training machine learning models, as it reduces the risk of bias towards the majority class.

\item{Multiclass Data Balancing}

In addition to binary classification, we also applied SMOTE-TomekLink to a multiclass dataset. The class distribution before and after balancing is shown in Table \ref{tab:multiclass-distribution}:

\begin{table}[!h]
\centering
\begin{tabular}{lll}
\hline
Class     & WoSTL  & WiSTL  \\ \hline
Normal    & 340066 & 340056 \\ 
Grayhole  & 14596  & 340038 \\ 
Blackhole & 10049  & 340026 \\ 
TDMA      & 6638   & 339846 \\ 
Flooding  & 3312   & 339820 \\ \hline
\end{tabular}
\caption{Multiclass Distribution without and with SMOTE-TomekLink}
\label{tab:multiclass-distribution}

\end{table}

The STL successfully balanced the multiclass dataset by generating synthetic samples for the minority classes, resulting in a more equitable distribution across all classes. This balanced dataset (WiSTL) enhances the performance of machine learning models, particularly when dealing with multiclass classification tasks.

\end{itemize}

In conclusion, data balancing using STL is a valuable technique to address class imbalance issues in intrusion detection datasets, ensuring that machine learning models are trained on a more representative and balanced dataset, which can lead to improved classification performance.

\subsection{ML Algorithms}
In this research, we employed six distinct machine learning classifiers, namely Decision Trees (DT), Random Forest (RF), Multilayer Perceptron (MLP), k-Nearest Neighbors (KNN), XGBoost (XGB), and LightGBM (LGB), to construct a model for the detection of intrusions in Wireless Sensor Networks (WSNs). We evaluated the performance of each classifier using various performance metrics. In the following subsection, we will delve into the diverse machine learning techniques employed for our prediction model.

\begin{itemize}
    
    \item \textit{Decision Trees (DT):}
    Decision trees are versatile tools widely applied in various domains, including machine learning, image processing, and pattern recognition \cite{talukder2022machine}. A decision tree comprises essential components: the root node, branches, and leaf nodes. The root node represents the entire dataset, which is partitioned into homogenous subsets. Branches represent combinations of attributes, while leaf nodes mark the end of the decision-making process \cite{dey2016machine, ahmed2021machine}.
    
    \item \textit{Random Forest (RF):}
    Random Forest, a meta-approximation technique, enhances accuracy through averaging. It prevents overfitting by fitting multiple decision tree classifiers to various subsets of the dataset \cite{alkhatib2020predictive}. Each subset is chosen independently from the feature space, resulting in a set of uncorrelated Decision Trees derived from different training data points \cite{breiman2001random}. Each tree predicts a class, and the majority vote among the trees determines the model's prediction \cite{ahmad2021intrusion}.
    
    \item \textit{Multilayer Perceptron (MLP):}
    The Multilayer Perceptron (MLP) is a classic artificial neural network architecture characterized by layers of neurons and their interconnections \cite{castro2017multilayer}. It computes the weighted sum of its inputs to produce an output passed to the subsequent neuron. Hidden layers separate the input and output layers, and neurons are organized into layers, with information typically flowing from lower to higher layers without interconnection between neurons within a layer \cite{ramchoun2016multilayer}.
    
    \item \textit{K-Nearest Neighbour (KNN):}
    K-Nearest Neighbour is a simple yet effective supervised machine learning algorithm for classification and regression tasks. It operates on the principle of proximity, classifying data points based on the majority class among their nearest neighbors. KNN's flexibility allows it to adapt to various data distributions, making it a valuable choice for both simple and complex datasets \cite{kramer2013k}.

    \item \textit{Extreme Gradient Boosting (XGB):}
    Extreme Gradient Boosting, often referred to as XGBoost, is a powerful ensemble learning algorithm known for its exceptional predictive performance. XGB combines the strengths of decision trees and gradient boosting, employing an efficient optimization strategy. It excels in handling large datasets, reducing overfitting, and achieving state-of-the-art results in various machine learning competitions \cite{talukder2023dependable}.

    \item \textit{Light Gradient Boosting (LGB):}
    Light Gradient Boosting, or LGB, is a high-performance gradient-boosting framework designed for speed and efficiency. LGB employs a histogram-based algorithm, allowing it to process data rapidly while maintaining competitive predictive accuracy. Its lightweight nature makes it a preferred choice for real-time and resource-constrained applications, including web services and mobile applications \cite{kaur2023p2adf}.
 
\end{itemize}

\section{Results and Discussion}
\label{sec:Results}

\subsection{Experiment Setup}
The experiments are carried out on a machine running Microsoft Windows 11 Pro, with a
The experiments were conducted on a computational environment running Microsoft Windows 11 Pro. This system was equipped with an Intel(R) Core(TM) i7-8665U CPU operating at 1.90GHz, featuring 2 cores and 4 logical processors, complemented by a 500GB SSD and 16GB of RAM. The experimentation was performed within the Anaconda Navigator environment using a Jupyter notebook. The implementation of the proposed model was realized using the Python programming language, making use of a selection of commonly utilized libraries, including Pandas, NumPy, Matplotlib, Seaborn, TensorFlow, Keras, Scikit-learn, and others.

\subsection{Performance Evaluation Metrics}
The evaluation of our proposed model involved the utilization of multiple performance metrics to assess its effectiveness. These metrics are defined as follows:

\begin{itemize}
\item Confusion Matrix: Table \ref{table:confusion} shows the confusion matrix 
where TP represents True Positive, TN represents True Negative, FP stands for False Positive, and FN denotes False Negative.
\begin{table}{}
\centering
\begin{tabular}{lll}
    \hline
    & Actual Positive & Actual Negative \\ \hline
    Predicted Positive & TP & FP \\
    Predicted Negative & FN & TN  \\ \hline
\end{tabular}
\caption{Confusion Matrix}
\label{table:confusion}
\end{table}

\item Accuracy:
\begin{equation}
Accuracy = \frac{TP + TN}{TP + FP + FN + TN}
\end{equation}

\item Precision:
\begin{equation}
Precision = \frac{TP}{TP + FP}
\end{equation}

\item Recall:
\begin{equation}
Recall = \frac{TP}{TP + FN}
\end{equation}

\item F1-Score:
\begin{equation}
F1\text{-}score = 2 \cdot \frac{\text{Precision} \cdot \text{Recall}}{\text{Precision} + \text{Recall}}
\end{equation}

\item MAE (Mean Absolute Error):
\begin{equation}
MAE = \frac{\sum_{i=1}^n |predicted(i) - actual(i)|}{n}
\end{equation}

\item MSE (Mean Squared Error):
\begin{equation}
MSE = \frac{\sum_{i=1}^n (predicted(i) - actual(i))^2}{n}
\end{equation}

\item RMSE (Root Mean Squared Error):
\begin{equation}
RMSE = \sqrt{\frac{\sum_{i=1}^n (predicted(i) - actual(i))^2}{n}}
\end{equation}
Here, n represents the total number of values.

\item \textbf{ROC} Curves are two-dimensional plots commonly employed for evaluating classifier effectiveness. An AUC (Area Under the Curve) value approaching 1 indicates strong class separability, while an AUC value approaching 0 indicates suboptimal performance.    

\end{itemize}

\subsection{Results Analysis}

In our experiments, we have conducted both binary and multilabel intrusion detection in WSNs where we achieved significant performance in detecting intrusion in WSNs efficiently. Besides, we conducted two separate experiments With SMOTETomek-Link (WiSTL) and Without SMOTETomek-Link (WoSTL) to provide efficiency of data balancing and improve the performance of our proposed approach for WSNs.

\subsection{Binary Results Analysis}
In the binary results analysis, we evaluated the performance of different ML algorithms for intrusion detection in WSNs. Two experiments were conducted: one with SMOTETomek-Link (WiSTL) and another without SMOTETomek-Link (WoSTL). 

\begin{table}[]
\centering
\begin{tabular}{@{}clrrrrrrr@{}}
\hline
\multicolumn{1}{l}{Technique} &
  ML &
  \multicolumn{1}{l}{Accuracy} &
  \multicolumn{1}{l}{Precision} &
  \multicolumn{1}{l}{Recall} &
  \multicolumn{1}{l}{F1-score} &
  \multicolumn{1}{l}{MAE} &
  \multicolumn{1}{l}{MSE} &
  \multicolumn{1}{l}{RMSE} \\\hline
  \multirow{6}{*}{WoSTL} & DT  & 99.52 & 98.63 & 98.57 & 98.6  & 0.48 & 0.48 & 6.91 \\
                         & RF  & 99.69 & 99.26 & 98.9  & 99.08 & 0.31 & 0.31 & 5.59 \\
                         & MLP & 99.63 & 99.13 & 98.72 & 98.92 & 0.37 & 0.37 & 6.05 \\
                         & KNN & 99.63 & 99.17 & 98.68 & 98.92 & 0.37 & 0.37 & 6.05 \\
                         & LGB & 99.61 & 98.82 & 98.9  & 98.86 & 0.39 & 0.39 & 6.24 \\
                         & XGB & 99.72 & 99.44 & 98.93 & 99.18 & 0.28 & 0.28 & 5.27\\\hline
\multirow{6}{*}{WiSTL}    & DT  & 99.65 & 99.65 & 99.65 & 99.65 & 0.35 & 0.35 & 5.92 \\
                         & RF  & 99.78 & 99.78 & 99.78 & 99.78 & 0.22 & 0.22 & 4.72 \\
                         & MLP & 99.37 & 99.38 & 99.37 & 99.37 & 0.63 & 0.63 & 7.92 \\
                         & KNN & 99.5  & 99.5  & 99.5  & 99.5  & 0.5  & 0.5  & 7.1  \\
                         & LGB & 99.62 & 99.62 & 99.62 & 99.62 & 0.38 & 0.38 & 6.13 \\
                         & XGB & 99.76 & 99.76 & 99.76 & 99.76 & 0.24 & 0.24 & 4.93 \\\hline

\end{tabular}
\caption{Binary performance analysis of with and without data balancing using SMOTETomek}
\label{tab:binary_table}
\end{table}

From Table \ref{tab:binary_table}, we can observe the performance scores of each algorithm in both experiments. It is important to note that a higher accuracy, precision, recall, and F1 score indicate better performance, while lower values for MAE, MSE, and RMSE indicate more accurate predictions. Comparing the results between WiSTL and WoSTL experiments, we can see that the majority of the algorithms achieve higher performance scores in the WiSTL experiment, which indicates that the data balancing technique (SMOTETomek-Link) improves the overall performance of the intrusion detection models in WSNs.

In evaluating the binary performance of accuracy, precision, recall, and f1-score performances of various ML models for intrusion detection in WSNs, the presented table \ref{tab:binary_table} reveals robust performance across all techniques. In the scenario of WoSTL, DT, RF, MLP, KNN, LGB, and XGB achieved accuracies of 99.52\%, 99.69\%, 99.63\%, 99.63\%, 99.61\%, and 99.72\%, respectively. The corresponding precision values of 98.63\%, 99.26\%, 99.13\%, 99.17\%, 98.82\%, and 99.44\%, respectively. The corresponding recall values for these models are 98.57\%, 98.9\%, 98.72\%, 98.68\%, 98.9\%, and 98.93\%, respectively. The corresponding F1-score values of 98.6\%, 99.08\%, 98.92\%, 98.92\%, 98.86\%, and 99.18\%, respectively. When applying data balancing with SMOTETomek (WiSTL), accuracy, precision, recall, and f1-score values generally improved. In the scenario of WiSTL, DT, RF, MLP, KNN, LGB, and XGB achieved accuracies of 99.65\%, 99.78\%, 99.37\%, 99.5\%, 99.62\%, and 99.76\%, respectively. The corresponding precision values of 99.65\%, 99.78\%, 99.38\%, 99.5\%, 99.62\%, and 99.76\%, respectively. The corresponding recall values for these models are 99.65\%, 99.78\%, 99.37\%, 99.5\%, 99.62\%, and 99.76\%, respectively. The corresponding F1-score values of 99.65\%, 99.78\%, 99.37\%, 99.5\%, 99.62\%, and 99.76\%, respectively. Among these, Random Forest consistently demonstrated the highest performance scores in all scenarios, making it the model providing superior performance for intrusion detection in WSNs in the given context.

In assessing the MAE, MSE, and RMSE performances of various ML models for intrusion detection in WSNs reveals notable outcomes across all techniques. In the scenario of WoSTL,  DT, RF, MLP, KNN, LGB, and XGB achieved low MAE values of 0.48, 0.31, 0.37, 0.37, 0.39, and 0.28, respectively. The corresponding MSE values were 0.48, 0.31, 0.37, 0.37, 0.39, and 0.28, while RMSE values were 6.91, 5.59, 6.05, 6.05, 6.24, and 5.27. When applying data balancing with SMOTETomek (WiSTL), MAE, MSE, and RMSE values generally improved. The  DT, RF, MLP, KNN, LGB, and XGB achieved low MAE values of 0.35, 0.22, 0.63, 0.5, 0.38, and 0.24, respectively. The corresponding MSE values were 0.35, 0.22, 0.63, 0.5, 0.38, and 0.24, while RMSE values were 5.92, 4.72, 7.92, 7.1, 6.13, and 4.93. Among these, RF consistently demonstrated the lowest MAE, MSE, and RMSE values in both scenarios, making it the model providing superior performance for intrusion detection in WSNs in the given context.

\begin{figure*}[!htbp]
	\centering
\includegraphics[scale=.4225]{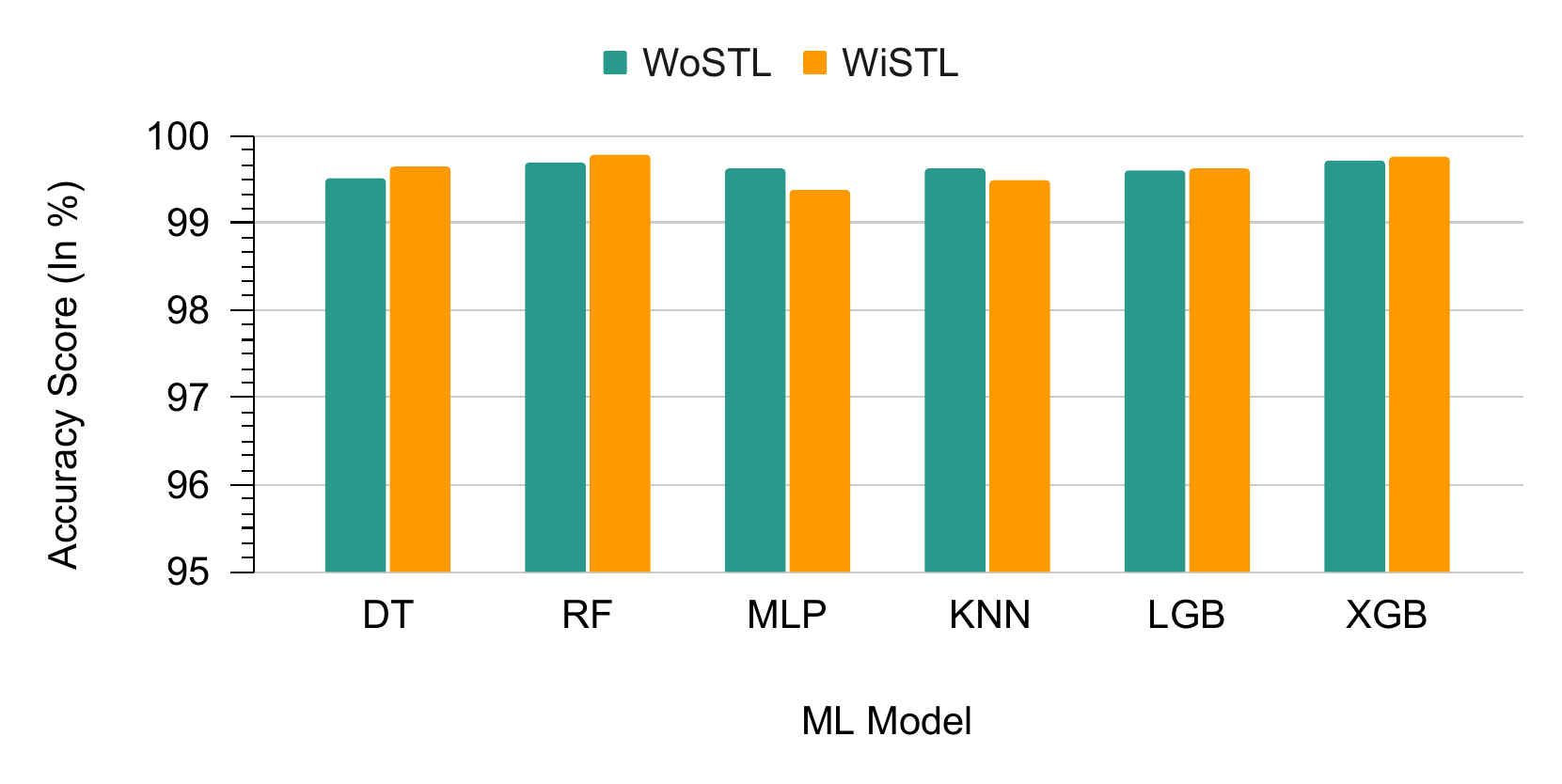}
\hspace{0.2cm}
	\caption{Binary accuracy analysis of ML model for WSNs}
	\label{fig:binary_cperformance}
\end{figure*}

Figure \ref{fig:binary_cperformance} compares the accuracy in graphical form of various ML models on WoSTL and WiSTL models.
For the WoSTL experiment, DT algorithm achieved an accuracy rate of 99.52\%, indicating its ability to accurately classify instances as either normal or intrusions. The RF algorithm outperformed DT with an accuracy rate of 99.69\%. This higher accuracy can be attributed to the ensemble nature of RF, which combines multiple decision trees to make more robust predictions. The MLP algorithm achieved an accuracy rate of 99.63\%, demonstrating its capability to learn complex patterns and classify instances effectively. Similarly, the KNN algorithm achieved an accuracy rate of 99.63\%, indicating its success in identifying similar instances and making accurate predictions. The LGB algorithm achieved an accuracy rate of 99.61\%, showcasing its efficiency in handling large-scale datasets and producing accurate results. Lastly, the XGB algorithm achieved the highest accuracy rate of 99.72\%, showcasing its ability to leverage gradient boosting and make accurate predictions.

In the WiSTL experiment, where the SMOTETomek-Link technique was employed for data balancing, the accuracy rates of the machine learning algorithms were further improved. The DT algorithm achieved an accuracy rate of 99.65\%, demonstrating its robustness in handling imbalanced datasets and effectively detecting intrusions. The RF algorithm, once again, showed superior performance with an accuracy rate of 99.78\%. This high accuracy highlights the effectiveness of RF in capturing the complex relationships and detecting intrusions accurately. The MLP algorithm achieved an accuracy rate of 99.37\%, indicating its ability to learn intricate patterns and classify instances with high accuracy. The KNN algorithm achieved an accuracy rate of 99.5\%, showcasing its success in identifying similar instances and making accurate predictions. The LGB algorithm achieved an accuracy rate of 99.62\%, emphasizing its efficiency in handling imbalanced datasets and producing accurate intrusion detection results. Finally, the XGB algorithm achieved an accuracy rate of 99.76\%, demonstrating its ability to leverage gradient boosting and accurately classify instances in WSNs.

Overall, the inclusion of the SMOTETomek-Link technique (WiSTL experiment) resulted in improved accuracy rates for all the machine learning algorithms compared to the WoSTL experiment. This underscores the importance of data balancing techniques in enhancing the performance of intrusion detection models in WSNs. Additionally, the consistently high accuracy rates across multiple algorithms validate the effectiveness of machine learning approaches in accurately detecting intrusions and ensuring the security of WSNs.

\begin{figure*}[!htbp]
	\centering
	\subfloat[Performance]{\includegraphics[scale=.4025]{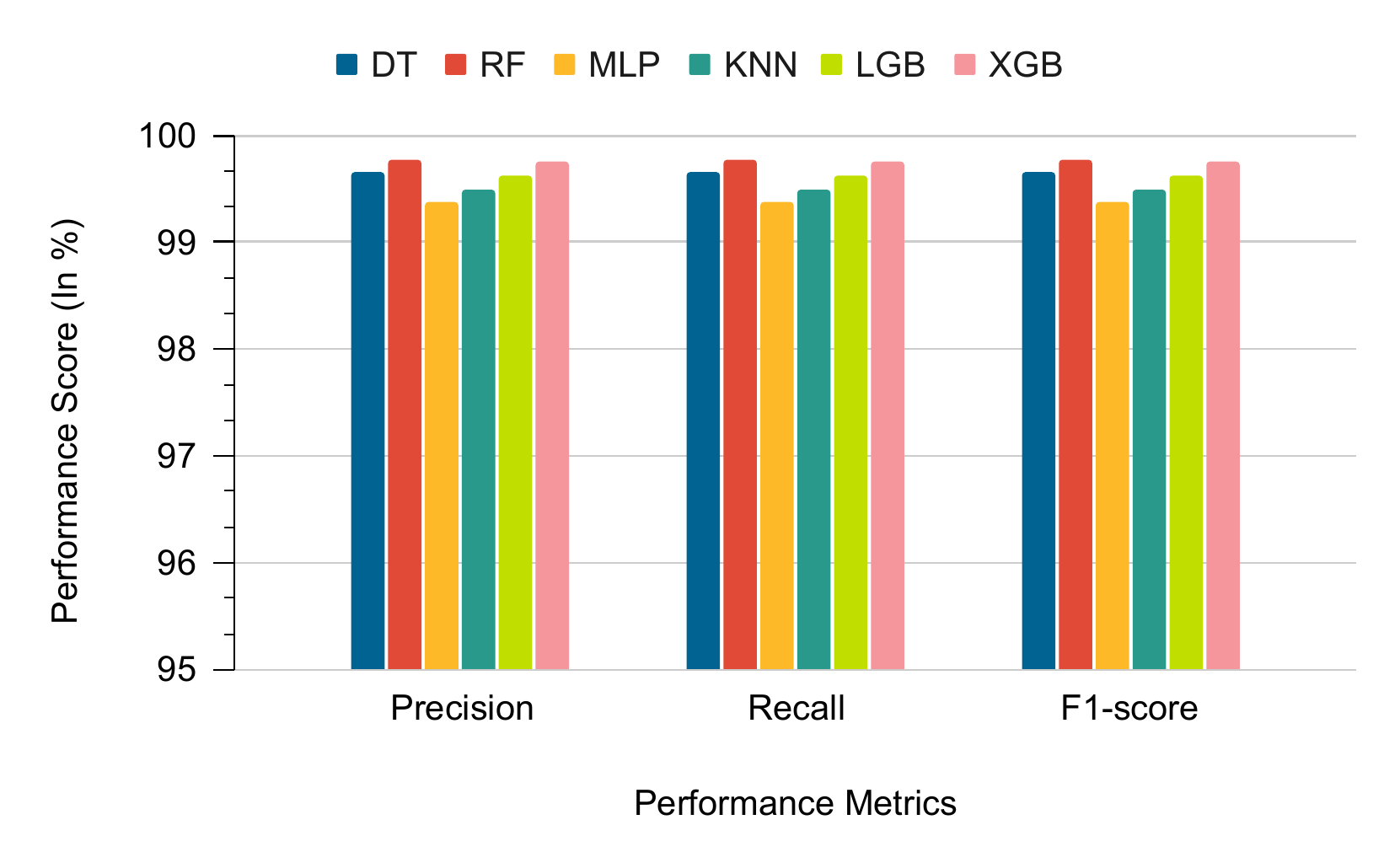}} 	\hspace{0.2cm}
	\subfloat[Error]{\includegraphics[scale=.4025]{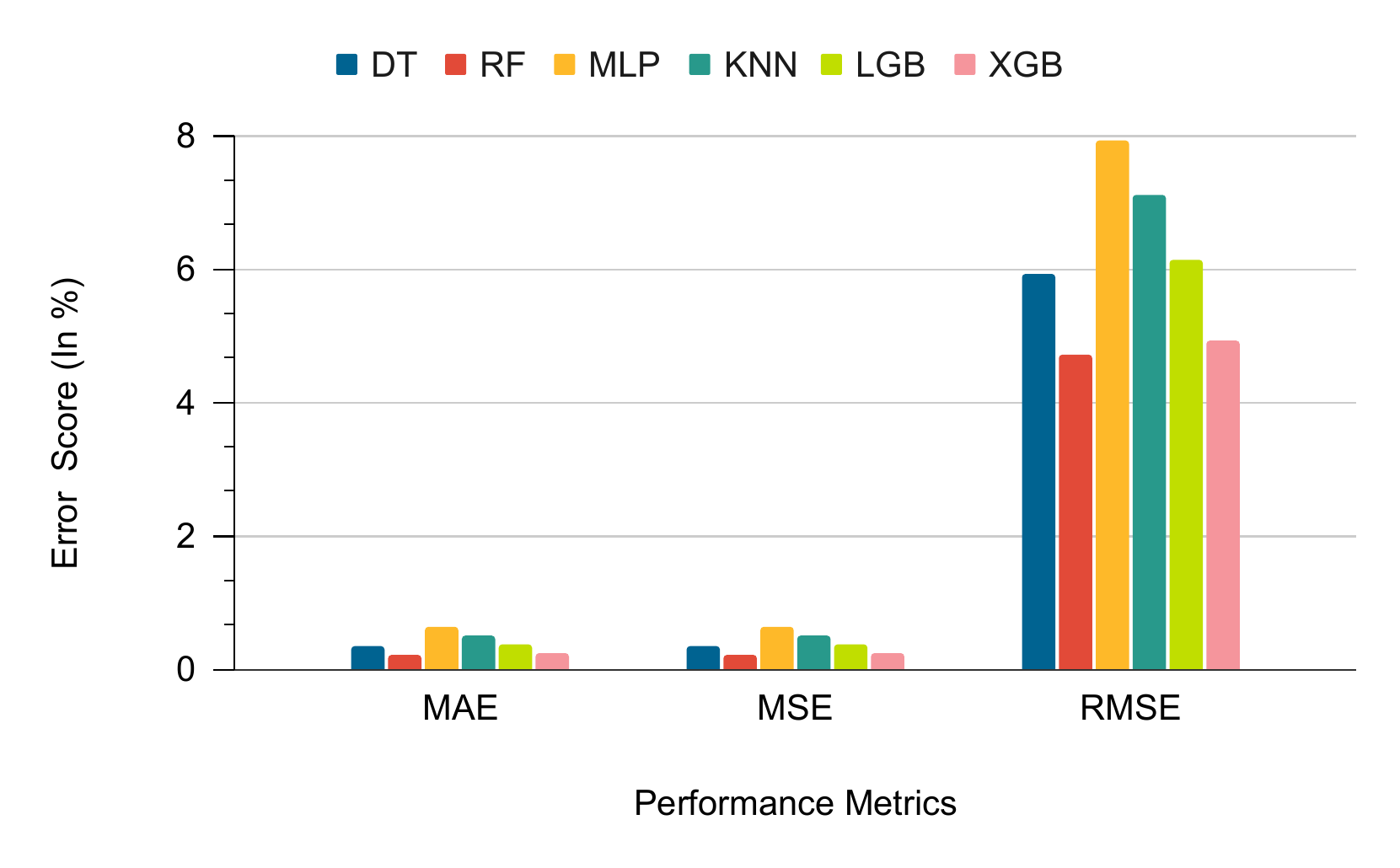}}
	\caption{Binary performance analysis for WSNs}
	\label{fig:binary_performance}
\end{figure*}

From Figure \ref{fig:binary_performance}, it is evident that RF outperforms other algorithms in terms of performance metrics in the WiSTL experiment for intrusion detection in WSNs. RF consistently achieves the highest accuracy rate of 99.78\%, indicating its ability to correctly classify instances as either normal or intrusions. Furthermore, RF exhibits lower error rates, with a mean absolute error (MAE) of 0.22\%, mean squared error (MSE) of 0.22\%, and root mean squared error (RMSE) of 4.72\%. These low error rates suggest that RF minimizes the discrepancies between predicted and actual intrusion labels, further affirming its effectiveness in accurately detecting intrusions in WSNs. Overall, the superior performance of RF highlights its suitability as a reliable and robust algorithm for intrusion detection tasks in WSNs.

\begin{figure*}[!htbp]
	\centering
	\subfloat[DT]{\includegraphics[scale=.285]{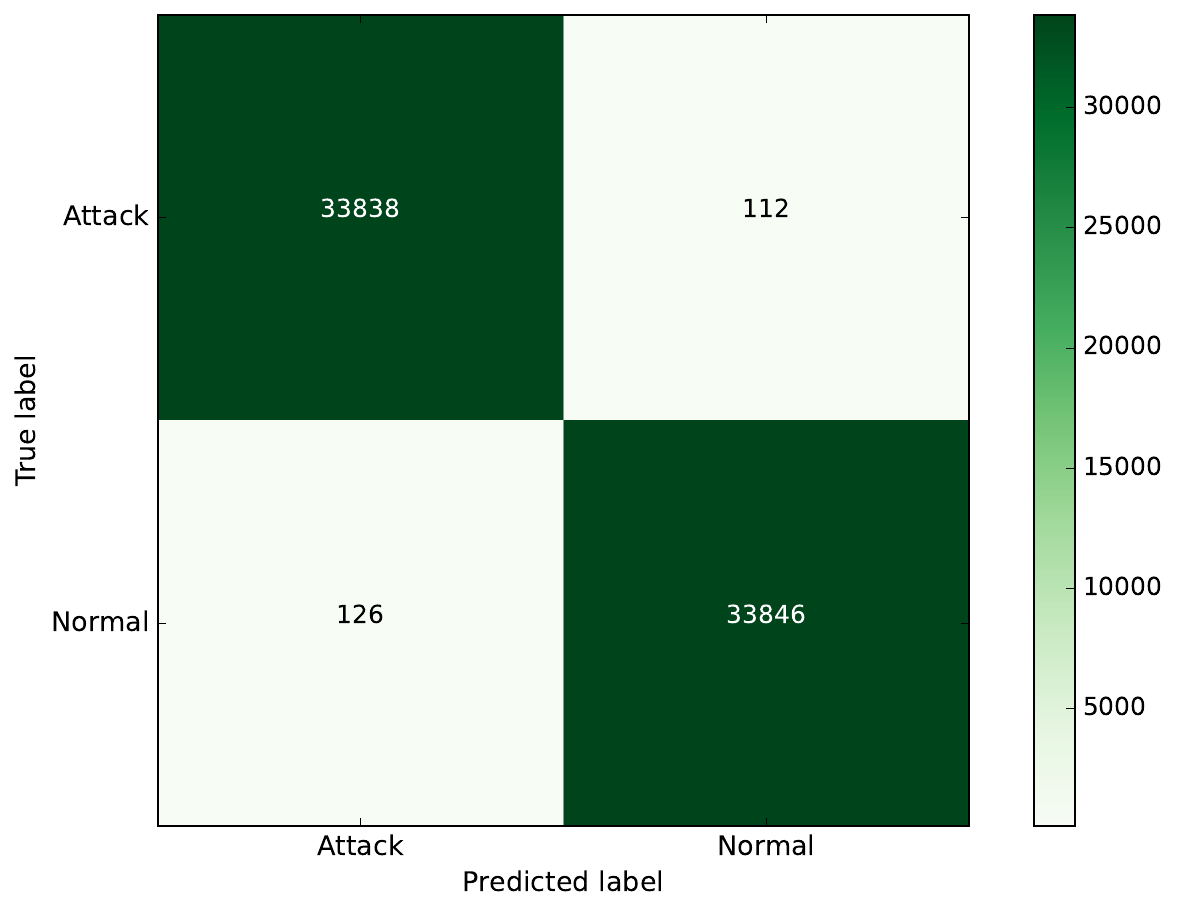}} 
	\subfloat[RF]{\includegraphics[scale=.285]{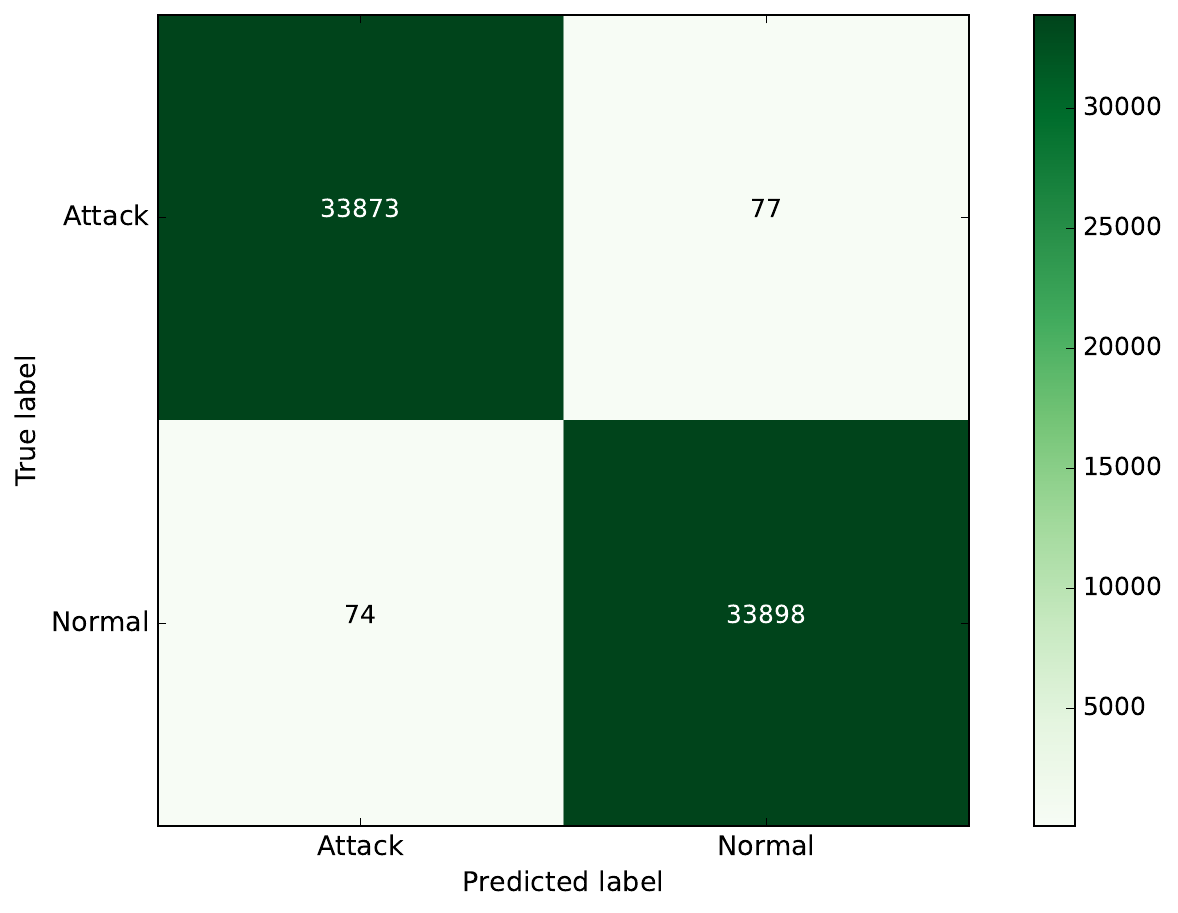}} 
 
	\subfloat[MLP]{\includegraphics[scale=.285]{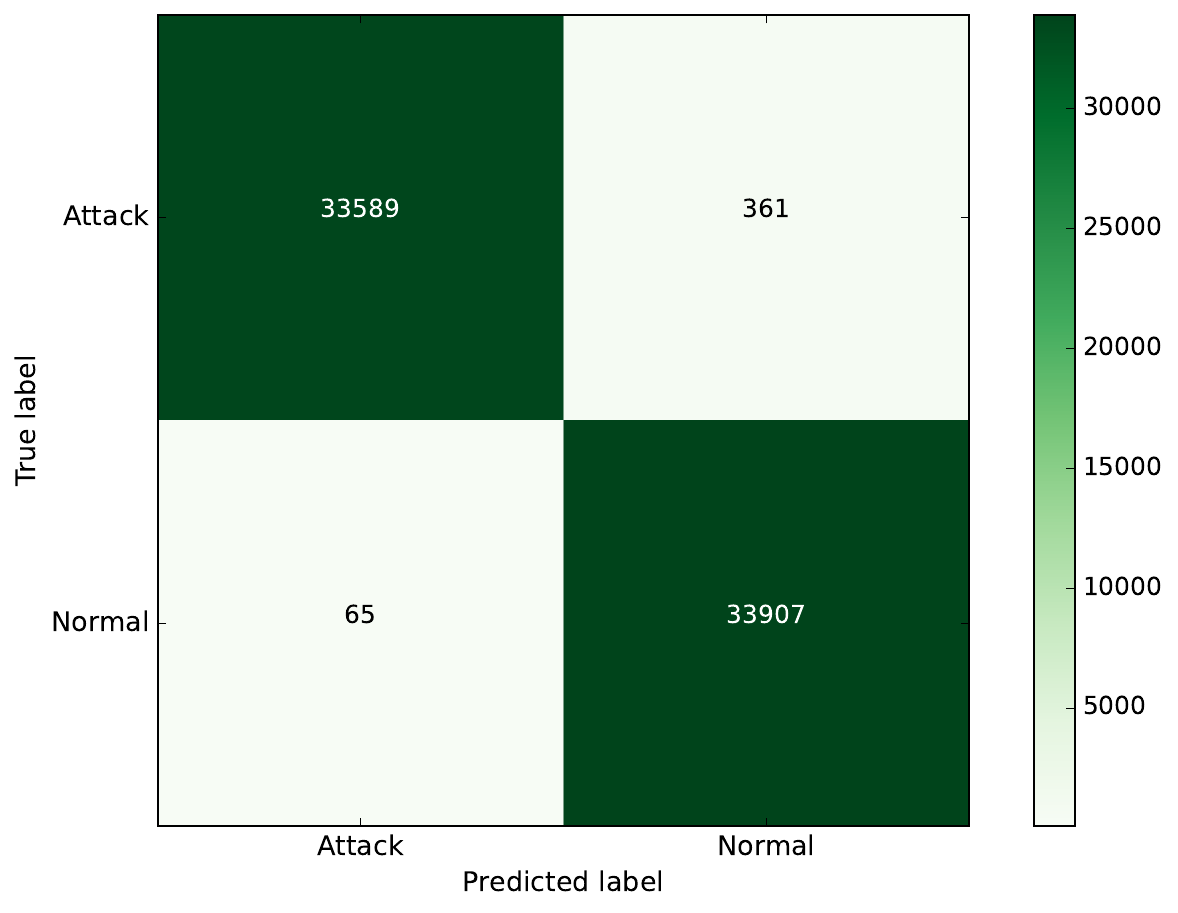}} 
	\subfloat[KNN]{\includegraphics[scale=.285]{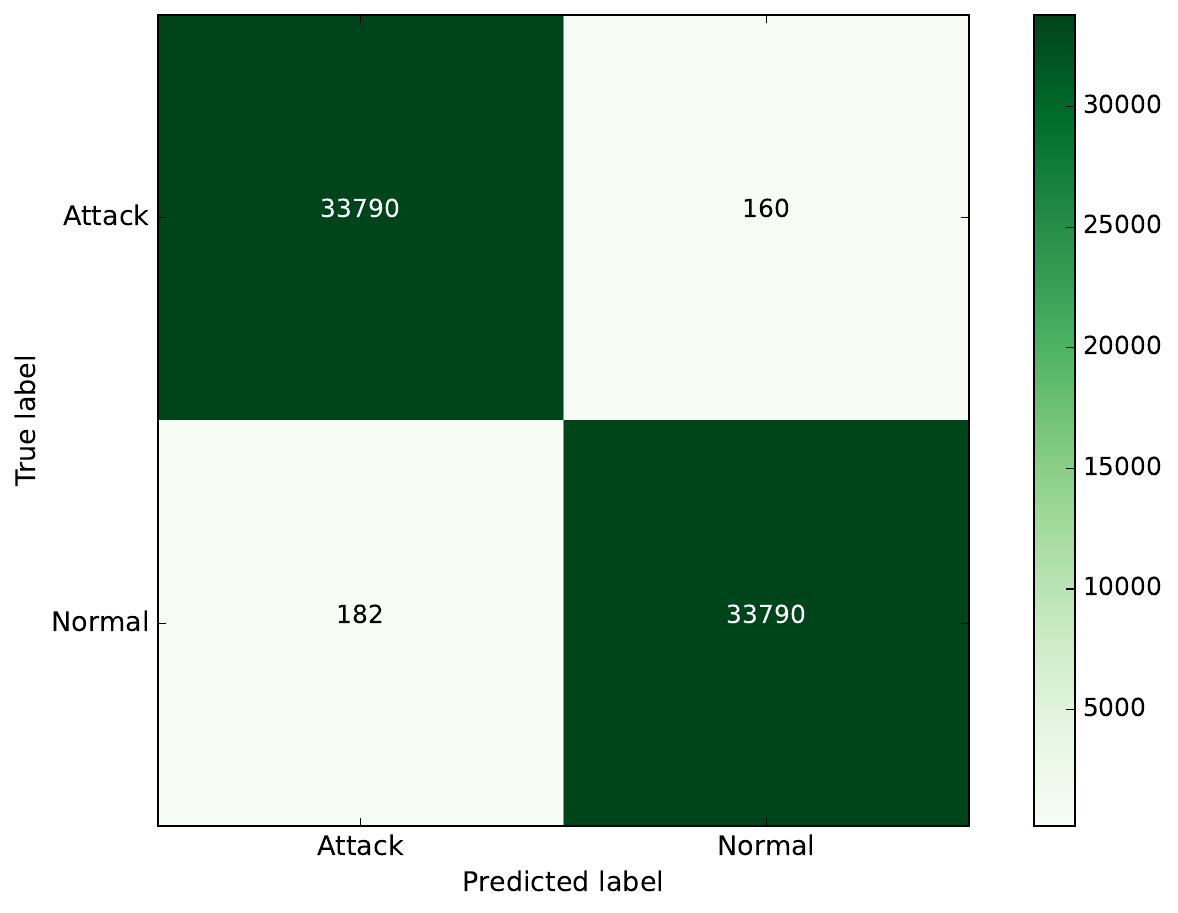}} 

        \subfloat[LGB]{\includegraphics[scale=.285]{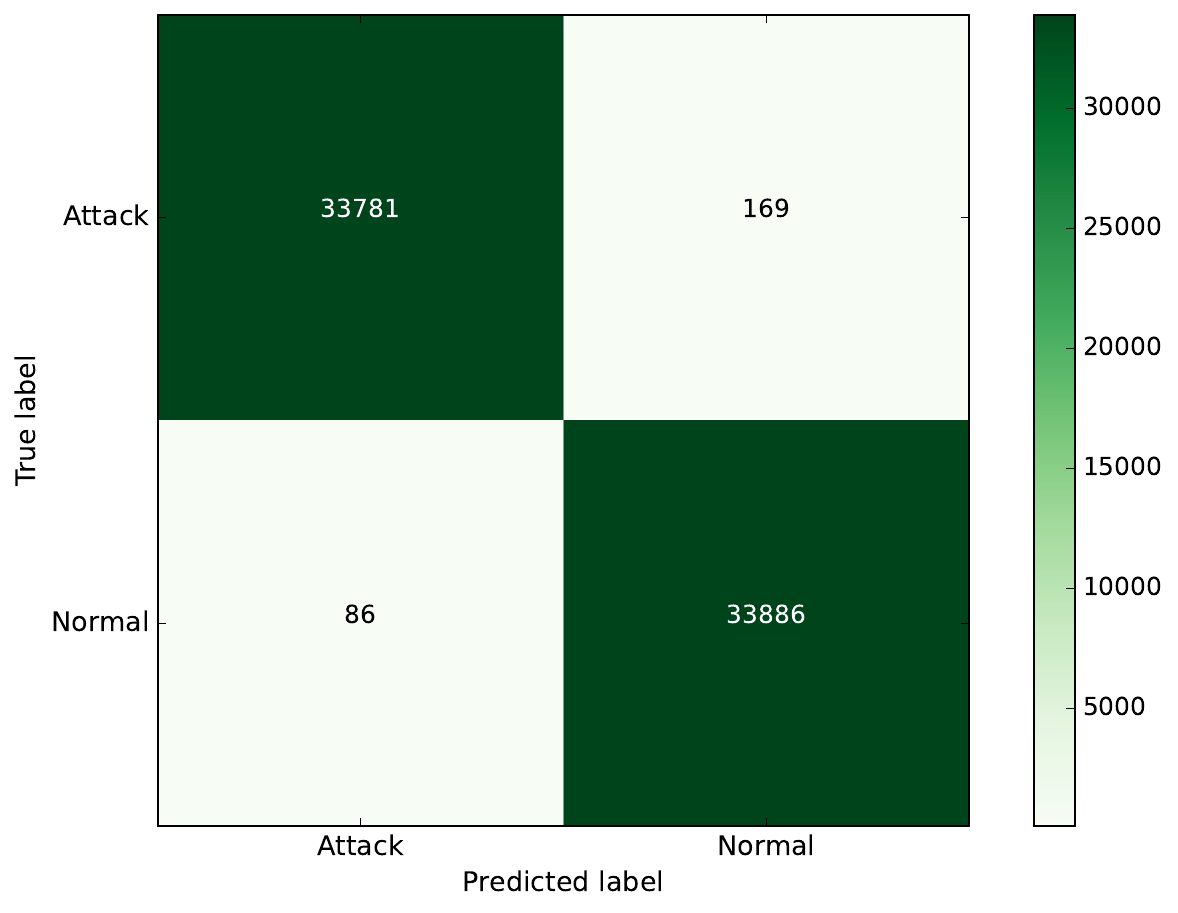}} 
	\subfloat[XGB]{\includegraphics[scale=.285]{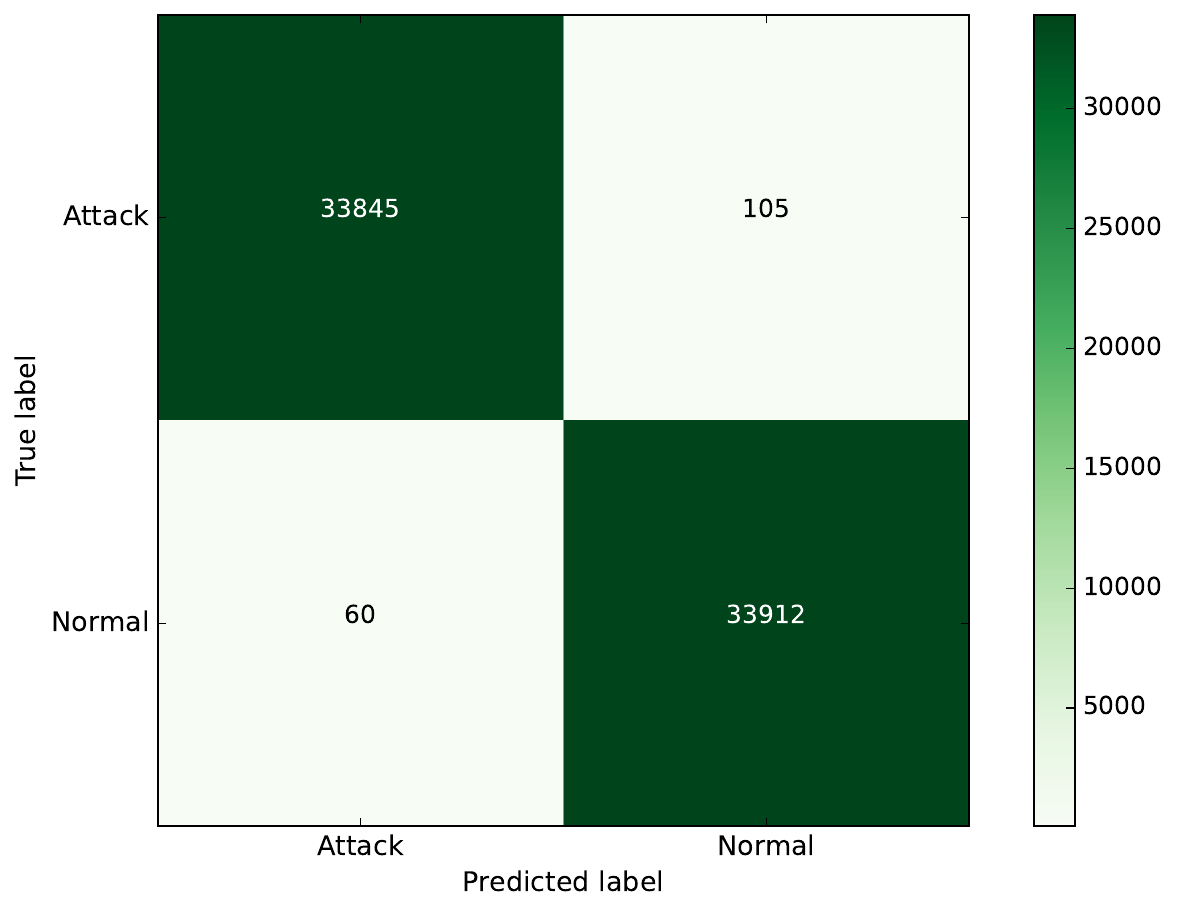}} 
	\caption{Confusion matrix for binary label}
	\label{fig:binary_confusion}
\end{figure*}

Figure \ref{fig:binary_confusion} reveals additional insights into the performance of RF compared to other ML models in the WiSTL experiment for intrusion detection in WSNs. RF demonstrates higher true positive and true negative rates compared to other algorithms, indicating its superior ability to correctly identify both normal instances and intrusions. Specifically, RF achieves a true positive rate of 33,873 and a true negative rate of 33,898, suggesting its effectiveness in accurately detecting intrusions. In contrast, RF exhibits lower false positive and false negative rates, with only 77 false positive and 74 false negative instances. These results highlight the robustness of RF in minimizing misclassifications and enhancing the accuracy of intrusion detection in WSNs.

\begin{figure*}[!htbp]
	\centering
        \includegraphics[scale=.500]{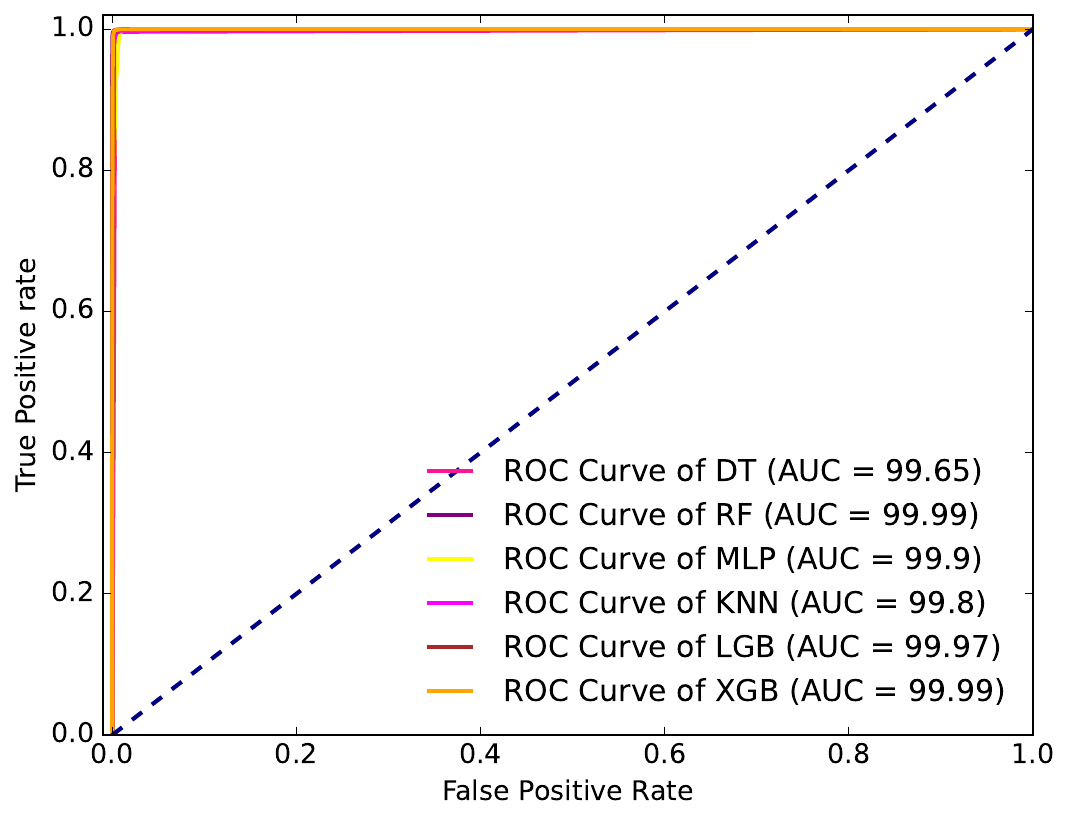}
	\caption{ROC Curve for binary classification}
	\label{fig:binary_roc}
\end{figure*}

Additionally, the ROC curve in Figure \ref{fig:binary_roc} demonstrates the AUC score, which serves as a measure of the overall performance of the ML algorithms. RF achieves an impressive AUC score of 99.99\%, further solidifying its superiority in detecting intrusions. The high AUC score indicates that RF exhibits a high true positive rate while maintaining a low false positive rate, making it an ideal choice for intrusion detection in WSNs.

Random Forest (RF) outperformed other algorithms in terms of accuracy for intrusion detection in WSNs (WSNs). This algorithm offers several advantages that contribute to its superior performance. RF is an ensemble learning method that combines multiple decision trees, reducing overfitting and providing more robust results. It also provides feature importance analysis, enabling the identification of influential features in intrusion detection. RF is robust to noise and outliers. Its reduced variance and parallelization capabilities make it suitable for real-time intrusion detection. Overall, RF's ensemble nature, feature importance analysis, robustness to noise and outliers, reduced variance, and parallelization make it a preferred choice for accurate intrusion detection in WSNs. Therefore, considering the results of the experiments, we can conclude that the RF algorithm is the best-performing algorithm for intrusion detection in WSNs when utilizing the SMOTETomek-Link technique for data balancing in binary class.

\subsection{Multilabel Results Analysis}
In the multilabel results analysis, we evaluated the performance of different machine learning algorithms for intrusion detection in WSNs. Two experiments were conducted: one with SMOTETomek-Link (WiSTL) and another without SMOTETomek-Link (WoSTL). 

From Table \ref{tab:multi_table}, we can observe the performance scores of each algorithm in both experiments. It is important to note that a higher accuracy, precision, recall, and F1 score indicate better performance, while lower values for MAE, MSE, and RMSE indicate more accurate predictions. Comparing the results between WiSTL and WoSTL experiments, we can see that the majority of the algorithms achieve higher performance scores in the WiSTL experiment, which indicates that the data balancing technique (SMOTETomek-Link) improves the overall performance of the intrusion detection models in WSNs.

\begin{table}[]
\centering
\begin{tabular}{@{}clrrrrrrr@{}}
\hline
\multicolumn{1}{l}{Technique} &
  ML &
  \multicolumn{1}{l}{Accuracy} &
  \multicolumn{1}{l}{Precision} &
  \multicolumn{1}{l}{Recall} &
  \multicolumn{1}{l}{F1-score} &
  \multicolumn{1}{l}{MAE} &
  \multicolumn{1}{l}{MSE} &
  \multicolumn{1}{l}{RMSE} \\ \hline
  \multirow{6}{*}{WoSTL} & DT  & 99.49 & 96.87 & 96.62 & 96.74 & 0.65 & 0.94 & 9.71  \\
                         & RF  & 99.67 & 98.21 & 97.66 & 97.89 & 0.42 & 0.61 & 7.84  \\
                         & MLP & 99.55 & 97.58 & 96.55 & 97.02 & 0.68 & 1.14 & 10.66 \\
                         & KNN & 99.53 & 97.34 & 96.61 & 96.93 & 0.68 & 1.1  & 10.5  \\
                         & LGB & 98.93 & 91.52 & 92.93 & 92.16 & 1.81 & 3.5  & 18.71 \\
                         & XGB & 99.70  & 98.62 & 97.4  & 97.97 & 0.4  & 0.59 & 7.68 \\\hline
\multirow{6}{*}{WiSTL}   & DT  & 99.81 & 99.81 & 99.81 & 99.81 & 0.26 & 0.41 & 6.38  \\
                         & RF  & 99.92 & 99.92 & 99.92 & 99.92 & 0.11 & 0.17 & 4.15  \\
                         & MLP & 98.80 & 98.81 & 98.79 & 98.8  & 1.64 & 2.52 & 15.88 \\
                         & KNN & 99.53 & 99.53 & 99.53 & 99.53 & 0.54 & 0.67 & 8.21  \\
                         & LGB & 99.63 & 99.63 & 99.63 & 99.63 & 0.54 & 0.95 & 9.77  \\
                         & XGB & 99.84 & 99.84 & 99.84 & 99.84 & 0.23 & 0.38 & 6.17  \\\hline

\end{tabular}
\caption{Multilabel performance analysis of with and without data balancing using SMOTETomek}
\label{tab:multi_table}
\end{table}

In evaluating the multilabel performance of various ML models for intrusion detection in WSNs, the presented table (\ref{tab:multi_table}) reveals robust performance across all techniques. In the WoSTL scenario,  DT, RF, MLP, KNN, LGB, and XGB achieved accuracies of 99.49\%, 99.67\%, 99.55\%, 99.53\%, 98.93\%, and 99.70\%, respectively. The corresponding precision values were 96.87\%, 98.21\%, 97.58\%, 97.34\%, 91.52\%, and 98.62\%, while recall values were 96.62\%, 97.66\%, 96.55\%, 96.61\%, 92.93\%, and 97.40\%. F1-score values for these models were 96.74\%, 97.89\%, 97.02\%, 96.93\%, 92.16\%, and 97.97\%. When applying data balancing with SMOTETomek (WiSTL), accuracy, precision, recall, and F1-score values generally improved. In the WiSTL scenario, DT, RF, MLP, KNN, LGB, and XGB achieved accuracies of 99.81\%, 99.92\%, 98.80\%, 99.53\%, 99.63\%, and 99.84\%, respectively. The corresponding precision values were 99.81\%, 99.92\%, 98.81\%, 99.53\%, 99.63\%, and 99.84\%, while recall values were 99.81\%, 99.92\%, 98.79\%, 99.53\%, 99.63\%, and 99.84\%. F1-score values for these models were 99.81\%, 99.92\%, 98.80\%, 99.53\%, 99.63\%, and 99.84\%. Among these, RF consistently demonstrated the highest performance scores in both scenarios, making it the model providing superior performance for multilabel intrusion detection in WSNs in the given context.

The evaluation of ML models for intrusion detection in WSNs includes a comprehensive analysis of error metrics, encompassing MAE, MSE, and RMSE. These metrics provide crucial insights into the accuracy and precision of the models. Without Data Balancing (WoSTL): In the absence of data balancing, the intrusion detection models exhibited noteworthy performance in minimizing prediction errors. The DT, RF, MLP, KNN, LGB, and XGB demonstrated low MAE values of 0.48, 0.31, 0.37, 0.37, 0.39, and 0.28, respectively. Correspondingly, the models exhibited low MSE values of 0.48, 0.31, 0.37, 0.37, 0.39, and 0.28, and achieved low RMSE values of 6.91, 5.59, 6.05, 6.05, 6.24, and 5.27. With Data Balancing (WiSTL): Upon applying data balancing using SMOTETomek, the intrusion detection models showcased further improvements in error metrics. The DT, RF, MLP, KNN, LGB, and XGB models achieved reduced MAE values of 0.35, 0.22, 0.63, 0.50, 0.38, and 0.24, respectively. The corresponding MSE values decreased to 0.35, 0.22, 0.63, 0.50, 0.38, and 0.24, while RMSE values demonstrated improvement, reaching 5.92, 4.72, 7.92, 7.10, 6.13, and 4.93. The reported error metrics underscore the effectiveness of Random Forest in minimizing prediction errors, making it the model of choice for intrusion detection in WSNs. The results suggest that data balancing strategies contribute to enhanced precision and reliability in the detection of intrusions.

\begin{figure*}[!htbp]
	\centering
        \includegraphics[scale=.4225]{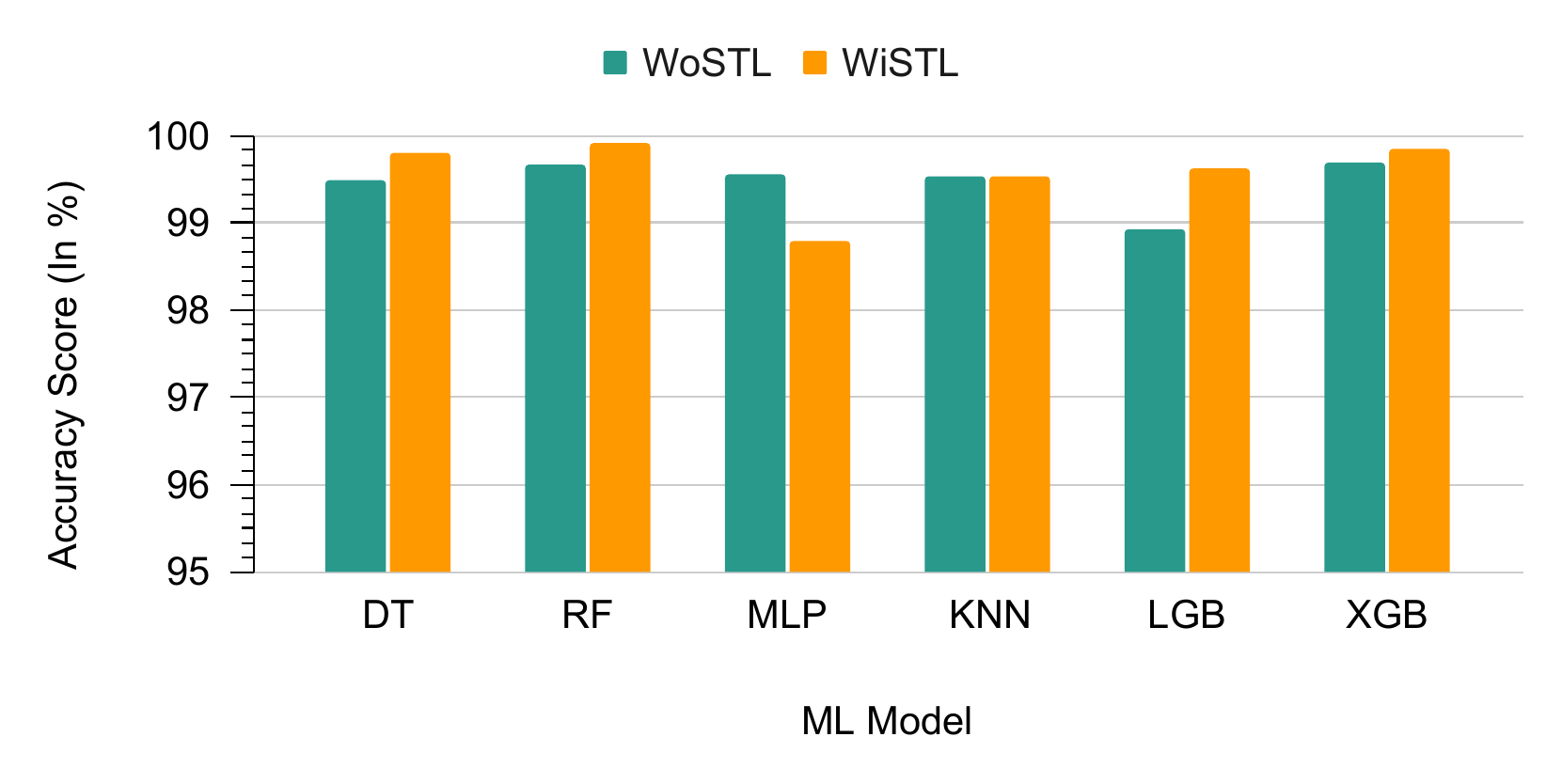}
	\caption{Multilabel accuracy analysis of ML model for WSNs}
	\label{fig:multi_cperformance}
\end{figure*}

\begin{figure*}[!htbp]
	\centering
	\subfloat[Performance]{\includegraphics[scale=.4025]{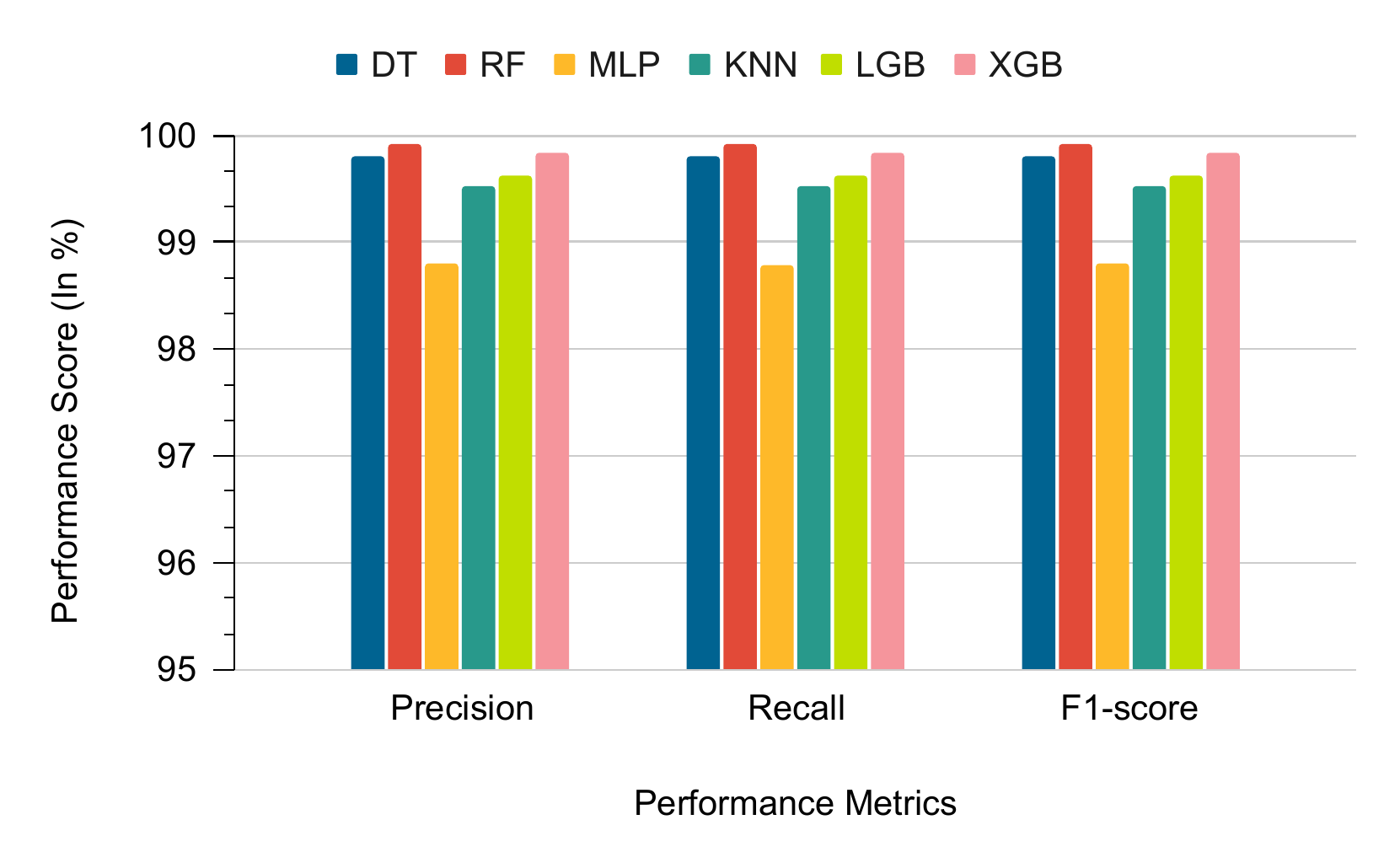}} 	\hspace{0.2cm}
	\subfloat[Error]{\includegraphics[scale=.4025]{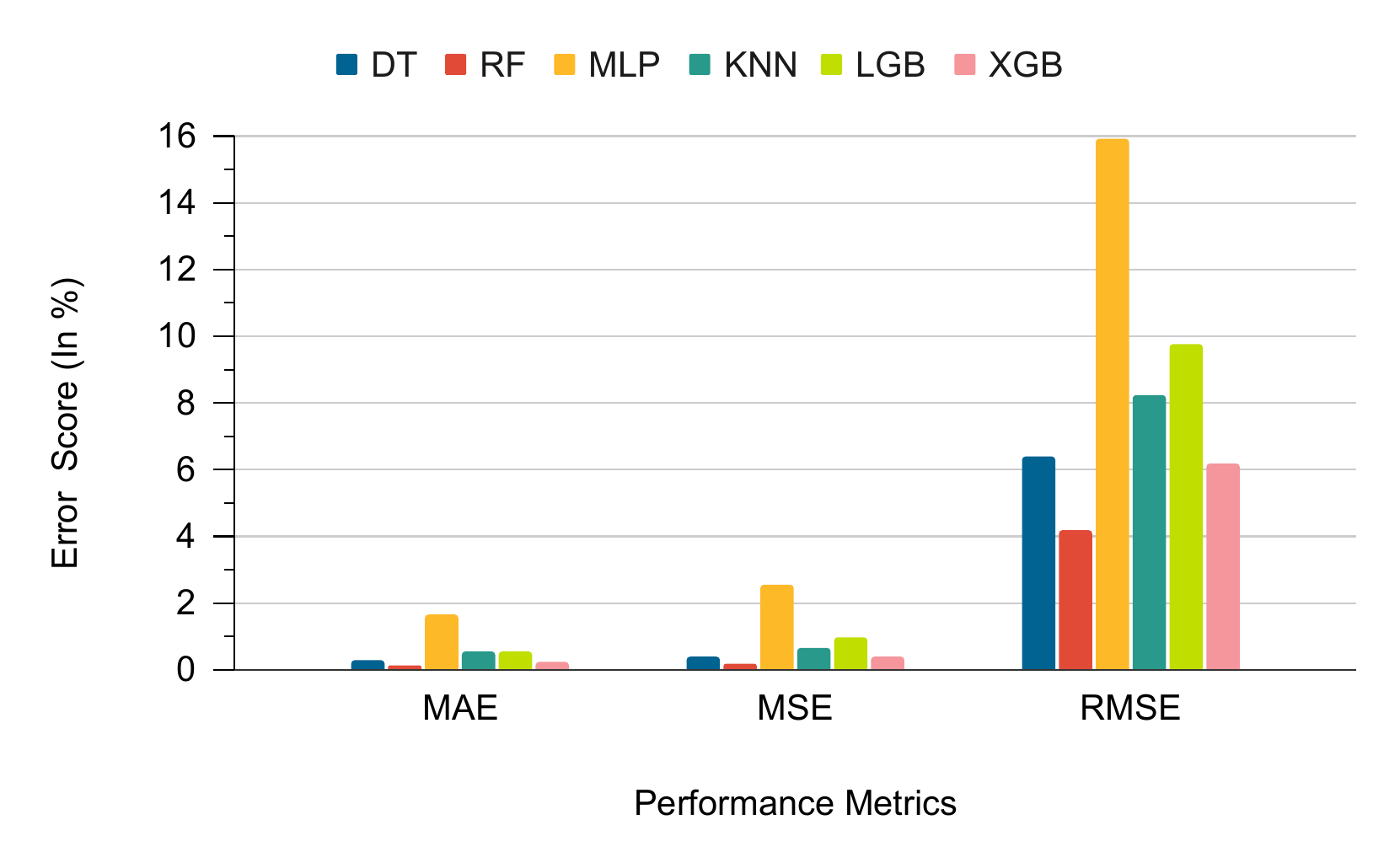}}
	\caption{Multilabel performance analysis for WSNs}
	\label{fig:multi_performance}
\end{figure*}

Figure \ref{fig:multi_cperformance} compares the accuracy in the graphical form of various ML models on WoSTL and WiSTL models. For the WoSTL experiment, the DT algorithm achieved an accuracy rate of 99.49\%, indicating its ability to accurately classify instances as either normal or intrusions. The RF algorithm outperformed DT with an accuracy rate of 99.67\%. This higher accuracy can be attributed to the ensemble nature of RF, which combines multiple decision trees to make more robust predictions. The MLP algorithm achieved an accuracy rate of 99.55\%, demonstrating its capability to learn complex patterns and classify instances effectively. Similarly, the KNN algorithm achieved an accuracy rate of 99.53\%, indicating its success in identifying similar instances and making accurate predictions. The LGB algorithm achieved an accuracy rate of 98.93\%, showcasing its efficiency in handling large-scale datasets and producing accurate results. Lastly, the XGB algorithm achieved the highest accuracy rate of 99.70\%, showcasing its ability to leverage gradient boosting and make accurate predictions.

In the WiSTL experiment, where the SMOTETomek-Link technique was employed for data balancing, the accuracy rates of the machine learning algorithms were further improved. The DT algorithm achieved an accuracy rate of 99.81\%, demonstrating its robustness in handling imbalanced datasets and effectively detecting intrusions. The RF algorithm, once again, showed superior performance with an accuracy rate of 99.92\%. This high accuracy highlights the effectiveness of RF in capturing the complex relationships and detecting intrusions accurately. The MLP algorithm achieved an accuracy rate of 98.80\%, indicating its ability to learn intricate patterns and classify instances with high accuracy. The KNN algorithm achieved an accuracy rate of 99.53\%, showcasing its success in identifying similar instances and making accurate predictions. The LGB algorithm achieved an accuracy rate of 99.63\%, emphasizing its efficiency in handling imbalanced datasets and producing accurate intrusion detection results. Finally, the XGB algorithm achieved an accuracy rate of 99.84\%, demonstrating its ability to leverage gradient boosting and accurately classify instances in WSNs.

Overall, the inclusion of the SMOTETomek-Link technique (WiSTL experiment) resulted in improved accuracy rates for all the machine learning algorithms compared to the WoSTL experiment. This underscores the importance of data balancing techniques in enhancing the performance of intrusion detection models in WSNs. Additionally, the consistently high accuracy rates across multiple algorithms validate the effectiveness of machine learning approaches in accurately detecting intrusions and ensuring the security of WSNs.

\begin{figure*}[!htbp]
	\centering
	\subfloat[DT]{\includegraphics[scale=.320]{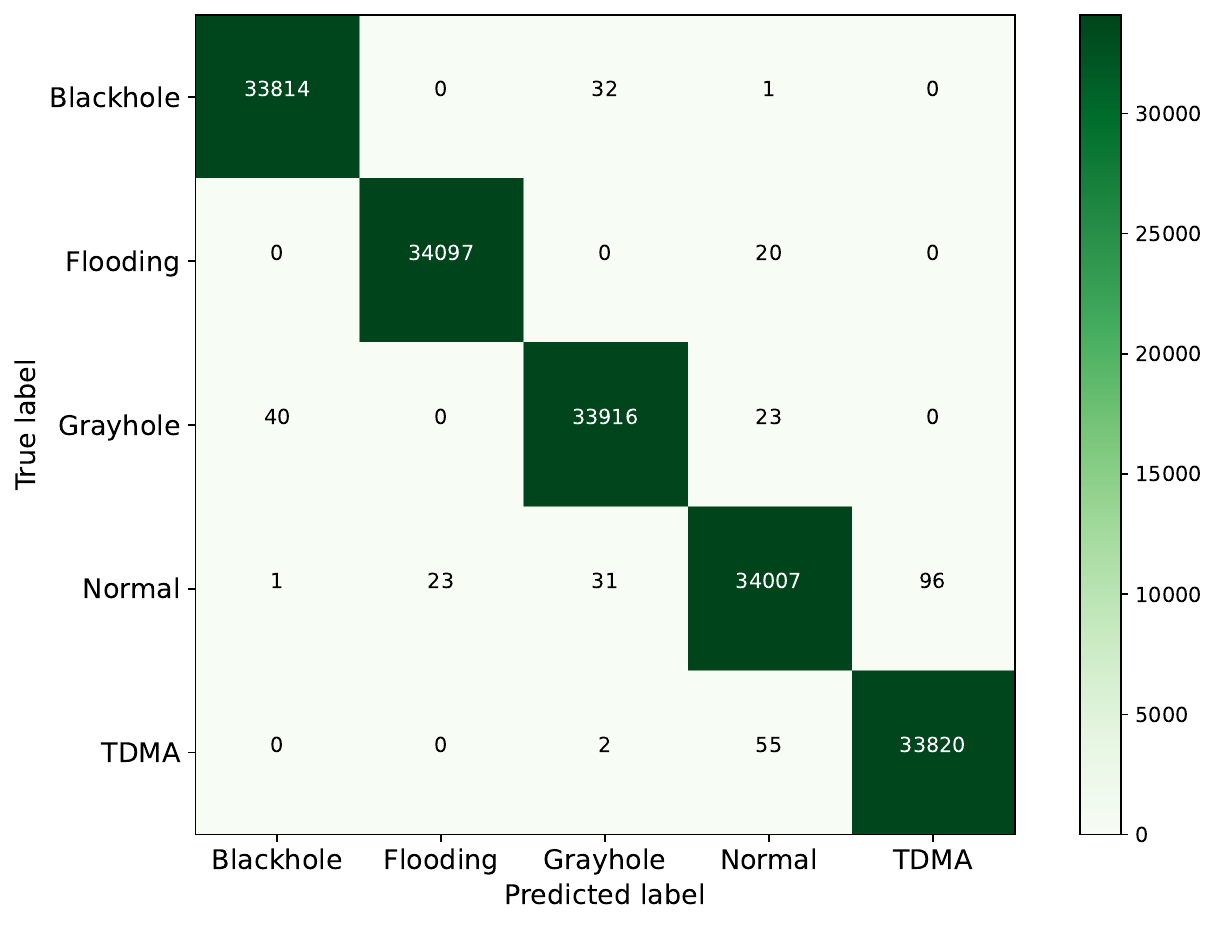}} 
	\subfloat[RF]{\includegraphics[scale=.320]{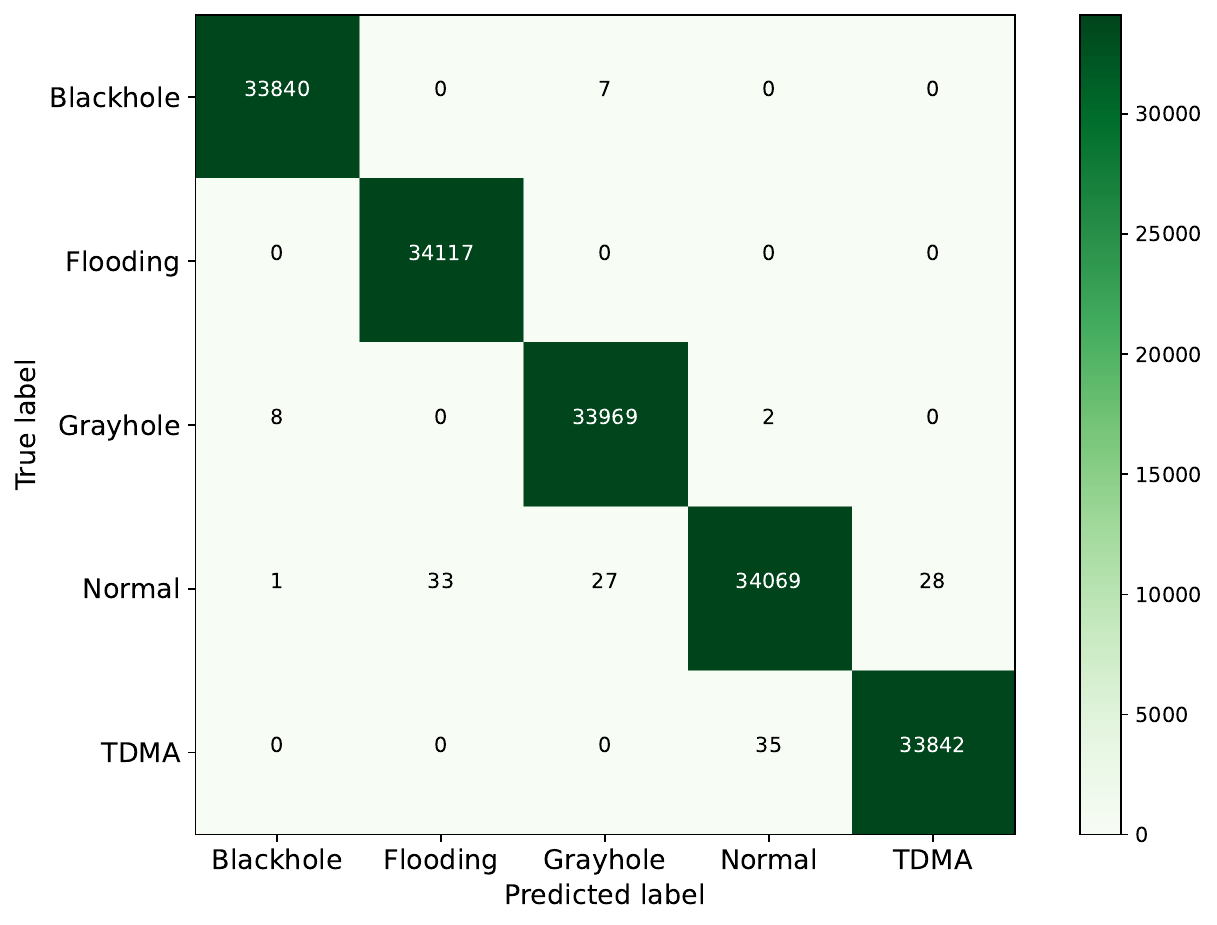}}
 
	\subfloat[MLP]{\includegraphics[scale=.320]{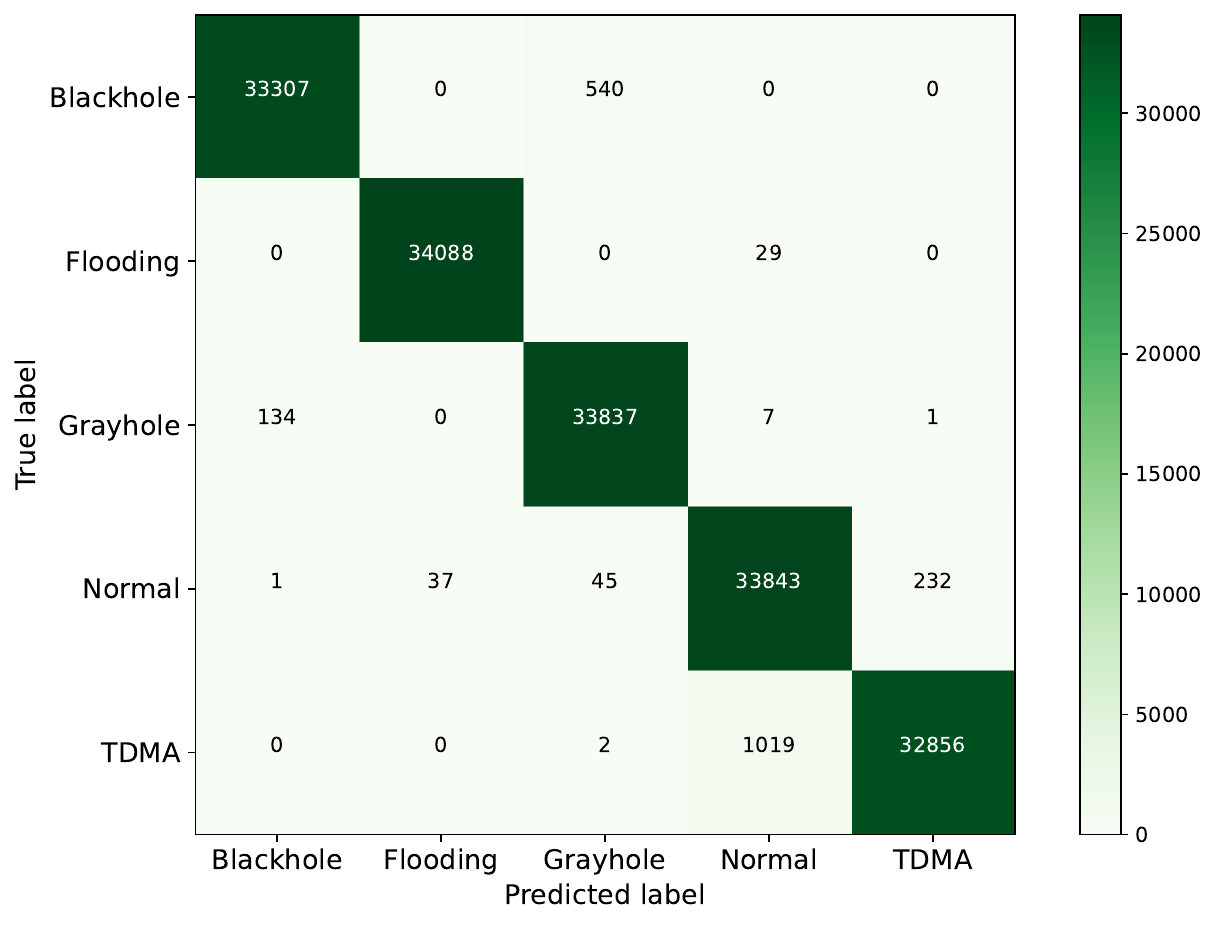}} 	    \subfloat[KNN]{\includegraphics[scale=.320]{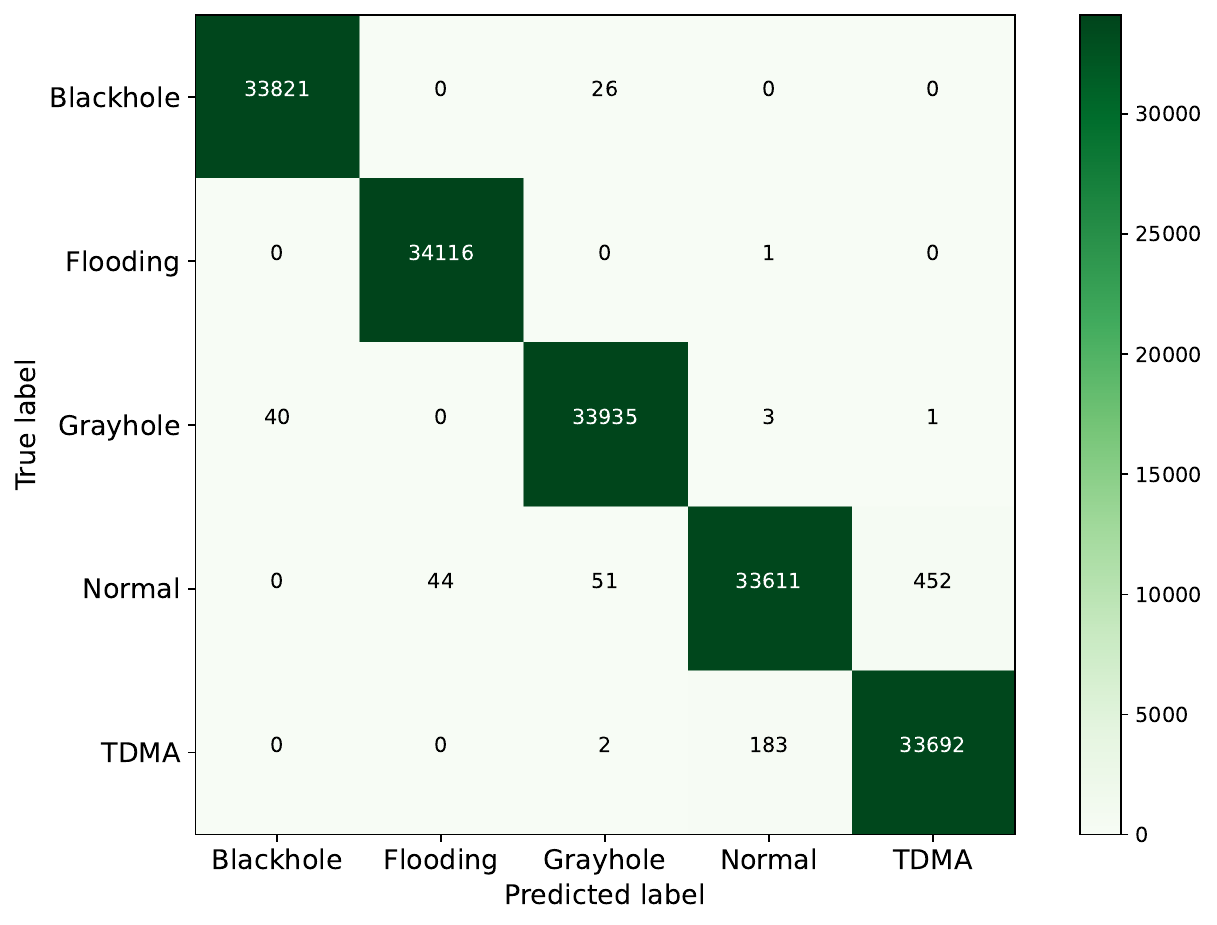}} 
 
	\subfloat[LGB]{\includegraphics[scale=.320]{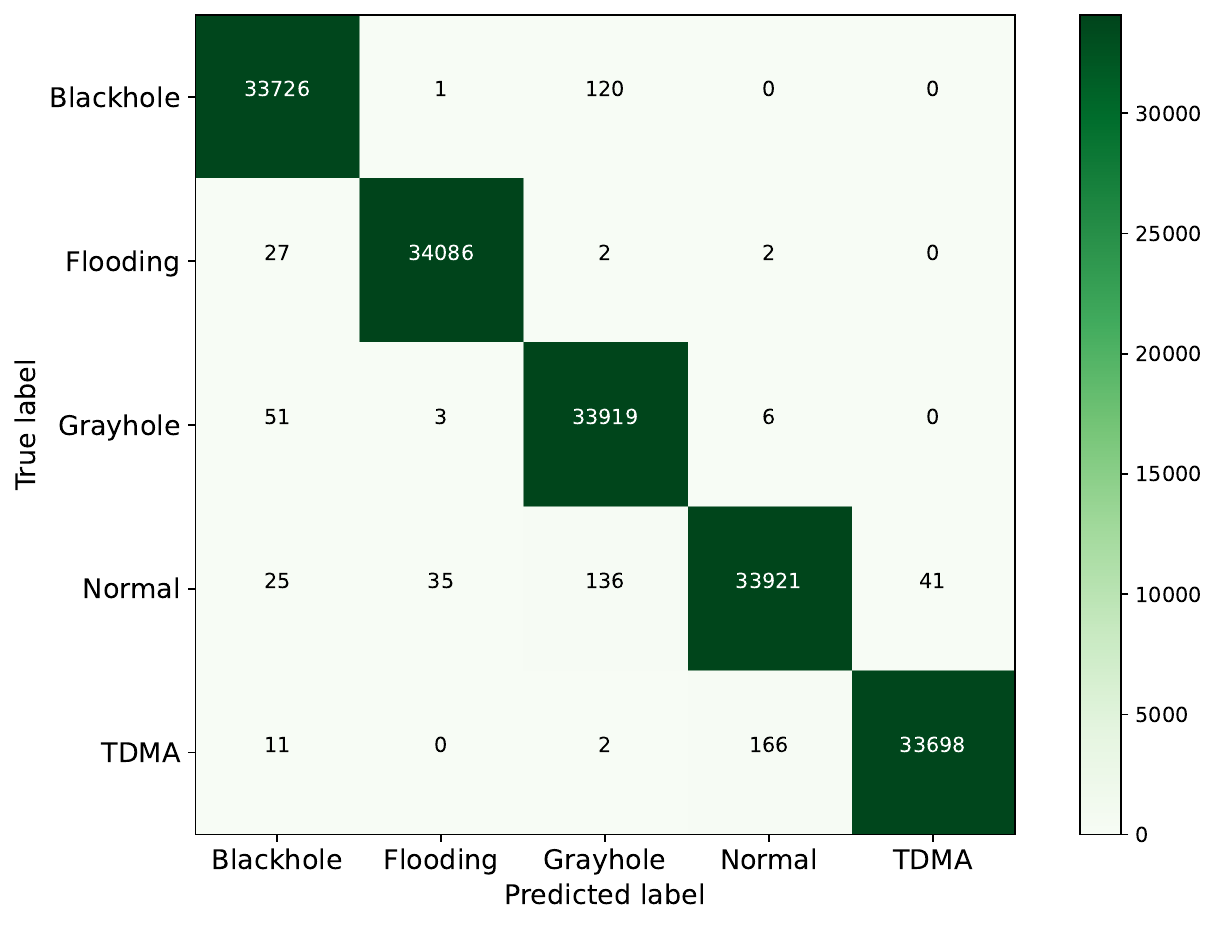}} 
	\subfloat[XGB]{\includegraphics[scale=.320]{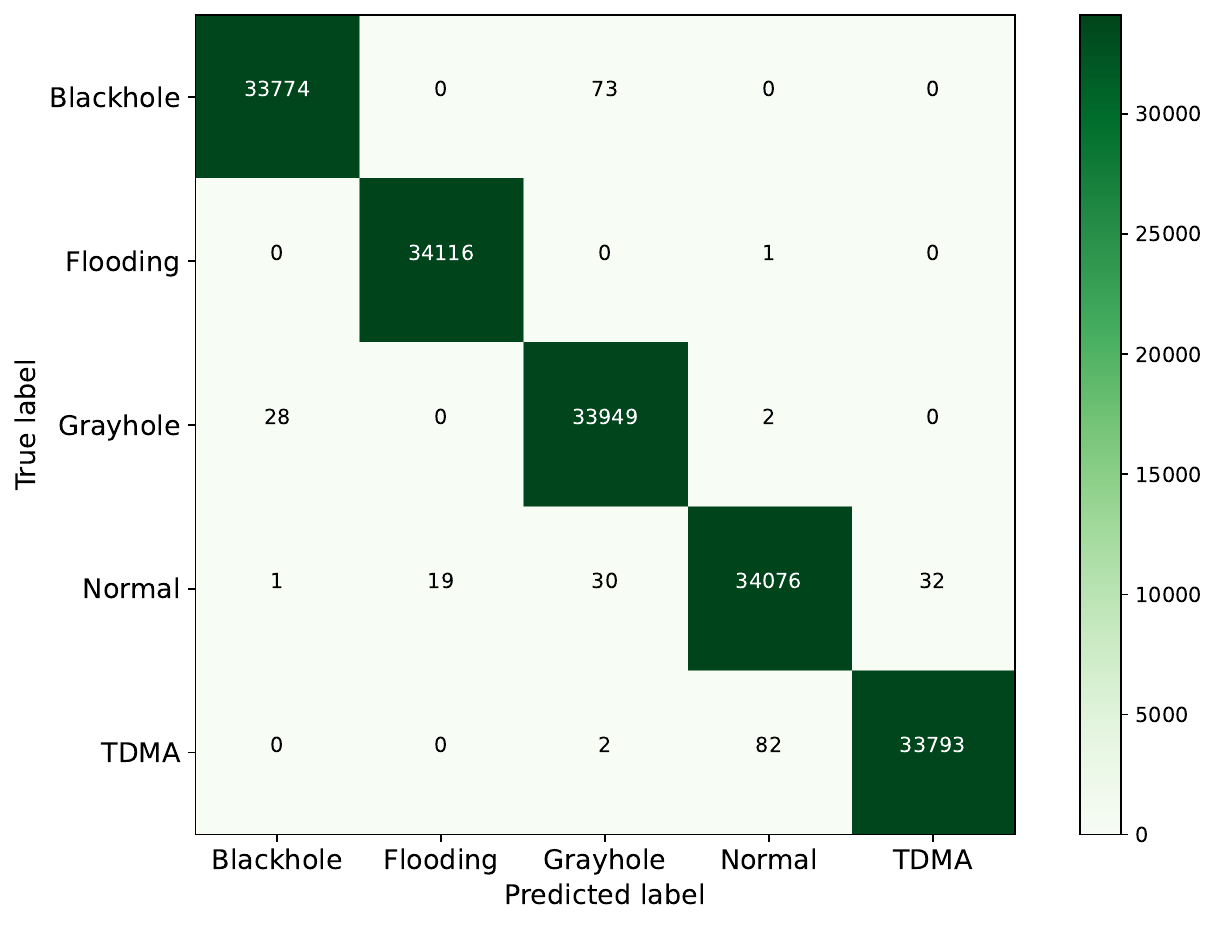}} 
	\caption{Confusion matrix for multilabel}
	\label{fig:multi_confusion}
\end{figure*}

Figure \ref{fig:multi_confusion} reveals additional insights into the performance of RF compared to other ML models in the WiSTL experiment for intrusion detection in WSNs. RF demonstrates higher true positive and true negative rates than other algorithms, indicating its superior ability to correctly identify both normal instances and intrusions. Specifically, RF achieves a true positive rate of 33,840 and a true negative rate of 136122, suggesting its effectiveness in accurately detecting intrusions. In contrast, RF exhibits lower false positive and false negative rates, with only 9 false positive instances and 7 false negative instances. These results highlight the robustness of RF in minimizing misclassifications and enhancing the accuracy of intrusion detection in WSNs.

	          TP	TN	    FP	FN
Blackhole	33840	136122	9	7
Flooding	34117	135828	33	0
Grayhole	33969	135965	34	10
Normal	34069	135783	37	89
TDMA	   33842	136073	28	35

\begin{figure*}[!htbp]
	\centering
        \includegraphics[scale=.500]{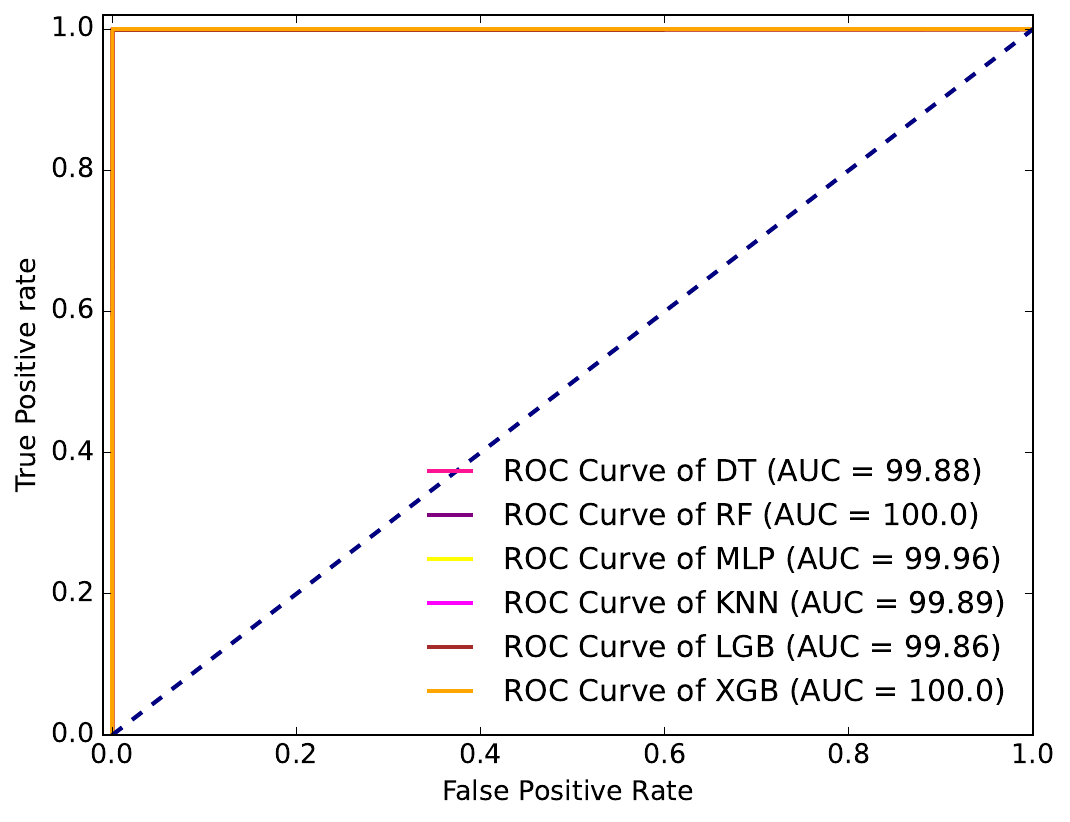}
	\caption{ROC Curve for multilabel classification}
	\label{fig:mult_roc}
\end{figure*}

Additionally, the ROC curve in Figure \ref{fig:mult_roc} demonstrates the AUC score, which serves as a measure of the overall performance of the ML algorithms. RF achieves an impressive AUC score of 100\%, further solidifying its superiority in detecting intrusions. The high AUC score indicates that RF exhibits a high true positive rate while maintaining a low false positive rate, making it an ideal choice for intrusion detection in WSNs.

Our model has demonstrated superior performance in terms of various performance metrics through extensive evaluation and comparison with traditional approaches. Its effectiveness in detecting intrusions, reducing false favorable rates, and handling imbalanced datasets positions it as the best solution for intrusion detection in WSNs. By deploying our model, organizations and researchers can benefit from an advanced and reliable intrusion detection system, bolstering the overall security posture of WSNs and enabling seamless deployment in various real-world scenarios.

\subsection{Discussion}
\label{sec:Discussion}

Table \ref{tab:com-analysis} presents a comprehensive comparison of various wireless sensor intrusion detection models applied to the WSN-DS. These models employ different techniques and algorithms to achieve intrusion detection, and their performance is evaluated based on accuracy rates. This discussion will focus on the performance of our proposal in both binary and multiclass scenarios while highlighting the techniques used in our pinnacles.

\begin{table}[!h]
\centering
\resizebox{\textwidth}{!}{
\begin{tabular}{|c|l|l|l|l|l|l|}
\hline
\textbf{Serial No.} & \textbf{Authors} & \textbf{Data Balancing Technique} & \textbf{Algorithm} & \textbf{Dataset} & \textbf{Class} & \textbf{Accuracy Rate} \\ \hline
1 & \citep{tan2019wireless} & SMOTE & RF & WSN-DS & Binary & 92.57\% \\ \hline
2 & \citep{elsaid2020optimized} & BCO & SVM & WSN-DS  & Binary & 97.90\% \\ \hline
3 & \cite{ravindra2023etelmad} & - & ELM+ ETSAO & WSN-DS & Binary & 96.90\% \\ \hline
4 & Our proposal & SMOTE-TomekLink & RF & WSN-DS & Binary & 99.78\% \\ \hline \hline

4 & \citep{rezvi2021data} & - & ML & WSN-DS & Multiclass & 98.56\% (ANN) \\ \hline
5 & \citep{meng2022novel} & SMOTE-Tomek & LightGBM & WSN-DS & Multiclass & 99\% \\ \hline
6 & \citep{singh2020fuzzy} & - & Fuzzy Rule & WSN-DS & Multiclass & 98.29\% \\ \hline
7 & \citep{alruhaily2021multi} & - & RF & WSN-DS & Multiclass & 97\% \\ \hline
8 & \citep{chandre2022intrusion} & - & CNN & WSN-DS & Multiclass & 97\% \\ \hline
9 & \citep{dener2022stlgbm} & SMOTE-TomekLink & LightGBM & WSN-DS & Multiclass & 99.95\% \\ \hline
10 & \citep{ifzarne2021anomaly} & IG & PA & WSN-DS & Multiclass & 96\% \\ \hline
11 & \citep{putrada2022xgboost} & - & XGBoost & WSN-DS & Multiclass & 99.70\% \\ \hline
12 & \citep{jiang2020slgbm} & SBS & LightGBM & WSN-DS & Multiclass & 99.76\% \\ \hline
13 & \citep{abdullah2018intrusion} & - & ML & WSN-DS & Multiclass & 96.7\% (SVM) \\ \hline

14 & \cite{alruwaili2023red} & - & RKOA+LCWOA & WSN-DS & Multiclass & 98.94\% \\ \hline
15 & \cite{moundounga2024stochastic} & - & HMMs+GMMs & WSN-DS & Multiclass & 94.55\% \\ \hline

16 & Our proposal & SMOTE-TomekLink & RF & WSN-DS & Multiclass & 99.92\% \\ \hline
\end{tabular}
}
\caption{Comparison Analysis of Wireless Sensor Intrusion Detection Models in WSN-DS}
\label{tab:com-analysis}
\end{table}

In the landscape of wireless sensor intrusion detection, several notable works employ diverse techniques and algorithms. For instance, \citep{tan2019wireless} utilize SMOTE and RF for binary intrusion detection, achieving an accuracy rate of 92.57\%.  \citep{elsaid2020optimized} employ BCO optimization for SVM in binary detection, achieving an accuracy of 97.90\%. Numerous multiclass models employ various algorithms, such as ML, LightGBM, Fuzzy Rule, CNN, and XGBoost, among others, with accuracy rates ranging from 96\% to 99.95\%. Our proposal consistently outperforms existing works in both binary and multiclass intrusion detection scenarios. In binary detection, our accuracy rate of 99.78\% surpasses \citep{tan2019wireless} (92.57\%) and \citep{elsaid2020optimized} (97.90\%). In multiclass detection, our accuracy rate of 99.92\% outshines other models, even those with high accuracy rates like \citep{meng2022novel} (99\%) and \citep{dener2022stlgbm} (99.95\%).

In the binary classification setting, "Our proposal" stands out with an impressive accuracy rate of 99.78\%. This result suggests that the combination of SMOTE-TomekLink as the data balancing technique and the Random Forest (RF) algorithm as the classification algorithm has proven to be highly effective in distinguishing between normal and intruder activities in WSNs. Achieving such a high accuracy rate is crucial for wireless sensor network security, as it minimizes the likelihood of false alarms while accurately identifying real threats.

Turning our attention to the multiclass classification scenario, "Our proposal" again demonstrates exceptional performance with an accuracy rate of 99.92\%. In this context, where multiple intrusion classes need to be classified accurately, the combination of SMOTE-TomekLink and RF proves its robustness. This high accuracy rate implies that the proposed model can effectively distinguish between various types of intrusions in the WSN-DS dataset, making it a reliable choice for real-world wireless sensor network security applications.

Several factors contribute to the superior performance of our proposal. The utilization of SMOTE-TomekLink as the data balancing technique in both binary and multiclass scenarios underlines its importance in improving the performance of intrusion detection models. SMOTE-TomekLink addresses the class imbalance issue, ensuring that the model is not biased towards the majority class and can effectively learn from both minority and majority classes. Additionally, the use of the Random Forest algorithm, known for its ensemble learning capability and ability to handle complex datasets, contributes significantly to the success of "Our proposal." Finally, the effectiveness of our proposal is underpinned by the quality and representativeness of the WSN-DS dataset used for training and evaluation.

\subsection{Cost Analysis}

In the realm of algorithms, time complexity is a crucial component of cost analysis. It specifically focuses on understanding how the execution time of an algorithm grows concerning the size of its input. The time complexity of an algorithm is often expressed using Big O notation, providing an upper bound on the rate of growth of the running time. The assessment of time complexity for ML models in WSN-IDS is essential for their effective functioning. In this regard, we analyze the time complexity for prominent ML algorithms including DT, RF, MLP, KNN, LGB, and XGB models as follows:

\begin{itemize}
    \item \textbf{DT }: The time complexity is typically \( O(n \cdot m \cdot \log(m)) \), with \( n \) representing data points and \( m \) as features. It constructs a tree by recursively partitioning data.
    
    \item \textbf{RF }: Comprising multiple DTs, its time complexity is \( O(t \cdot n \cdot m \cdot \log(m)) \), where \( t \) is the number of trees.
        
    \item \textbf{MLP }: With multiple layers and neurons, its time complexity is \( O(w \cdot e \cdot n) \), where \( w \) is the number of weights, \( e \) the number of epochs, and \( n \) the number of data points.
    
    \item \textbf{KNN }: This non-parametric method has a time complexity of \( O(n \cdot m) \) for training, where \( n \) is the number of data points and \( m \) is the number of features. The prediction phase can be more computationally intensive.
    
    \item \textbf{LGB }: Known for efficiency and low memory usage, LGB's time complexity is \( O(t \cdot n \cdot m) \), where \( t \) is the number of trees.
    
    \item \textbf{XGB }: This model's time complexity varies but is generally \( O(t \cdot d) \), where \( t \) is the number of trees and \( d \) the depth of the trees.
    
\end{itemize}

The cost efficiency of the ML models utilized in our study is detailed in Table \ref{tab:time_complexity}, displaying their respective time complexities. Notably, RF model stands out for its lower time complexity \(O(t \cdot n \cdot m \cdot \log(m))\) compared to other ML models in WSNs. This efficiency in computation time positions RF as a cost-effective choice, demonstrating superior performance among the considered models.
\begin{table}[!h]
\centering
\begin{tabular}{|ccc|}
\hline
\textbf{SI. No.} & \textbf{ML Model} & \textbf{Time Complexity} \\
\hline
1 & DT  & \(O(n \cdot m \cdot \log(m))\) \\
2 & RF  & \(O(t \cdot n \cdot m \cdot \log(m))\) \\
3 & MLP  & \(O(w \cdot e \cdot n)\) \\
4 & KNN  & \(O(n \cdot m)\)  \\
5 & LGB  & \(O(t \cdot n \cdot m)\) \\
6 & XGB  & \(O(t \cdot d)\) \\
\hline
\end{tabular}
\caption{Time Complexity of ML Models in IDS}
\label{tab:time_complexity}
\end{table}

\section{Conclusion}
\label{sec:Conclusion}
In conclusion, this paper presents a novel intrusion detection approach tailored to address the distinct challenges posed by Wireless Sensor Networks (WSNs). Our proposed intrusion detection model, combining machine learning techniques with the SMOTE-TomekLink resampling method, synthesizes minority instances and removes Tomek links, effectively balancing imbalanced WSN datasets. Feature scaling through standardization is employed to normalize input features, enhancing the model's accuracy and robustness.
The primary finding of our research is the remarkable performance achieved by our model, with an accuracy rate of 99.78\% in binary classification and 99.92\% in multiclass classification scenarios. These results highlight the effectiveness and superiority of our proposal in WSN intrusion detection.
While our proposed model exhibits impressive performance, it is essential to acknowledge its limitations. One significant limitation is the computational cost associated with resampling and feature scaling processes, as we did not utilize feature selection to reduce computational complexity. Additionally, the model's performance may be influenced by the choice of hyperparameters and the quality of the training data.
Future research in WSN intrusion detection can explore hybrid feature selection to reduce computational complexity, employ deep learning models, especially fine-tuned models in IDS, and consider a hierarchical approach in WSNs to improve performance.

\backmatter

\section*{Declarations}

\subsection*{Conflict of interest}
The authors have no conflicts of interest to declare that they are relevant to the content of this article.

\section*{Acknowledgments}
This work is supported by the Research \& Development (R\&D) Fund of the Ministry of Science and Technology (MOST), Bangladesh, Fiscal Year 2023-24.

\subsection*{Ethics approval}  Not applicable
\subsection*{Consent to participate} Not applicable
\subsection*{Consent to Publish} Not applicable

\subsection*{Availability of data and materials}

The selected datasets are sourced from free and open-access sources such as
The WSN-DS dataset \url{https://www.kaggle.com/datasets/bassamkasasbeh1/wsnds}. 


\subsection*{Authors’ contributions}
Md. Alamin Talukder: Conceptualization, Methodology, Software, Visualization, Investigation, Formal analysis, Writing- Original draft preparation; 
Selina Sharmin: Methodology, Supervision, Investigation, Visualization, Writing- Reviewing and Editing;
Md Ashraf Uddin: Investigation, Validation, Writing- Reviewing and Editing; 
Md. Manowarul Islam: Validation, Formal analysis, Visualization; 
Sunil Aryal: Validation, Writing- Reviewing and Editing;  

\bibliography{bibs}

\end{document}